\documentclass[cernpreprint,atlasdraft=true,texlive=2016,UKenglish,texmf,texmf]{atlasdoc}
 
\usepackage{atlaspackage}
\usepackage{atlasbiblatex}
 
\usepackage{atlasphysics}
 
\graphicspath{{logos/}{figures/}}
 
\usepackage{ANA-TOPQ-2018-05-PAPER-defs}

\AtlasTitle{Evidence for \tttt{} production in the multilepton final state in proton--proton collisions at $\sqrt{s}=13$~\tev\ with the ATLAS detector}
 
\AtlasAbstract{
A search is presented for four-top-quark production using an integrated luminosity of 139~fb\(^{-1}\)
of proton--proton collision data at a centre-of-mass energy of $13~\tev$ collected by
the ATLAS detector at the LHC.
Events are selected if they contain a same-sign lepton pair
or at least three leptons (electrons or muons).
Jet multiplicity, jet flavour and event kinematics are used to separate signal
from the background through a multivariate discriminant,
and dedicated control regions are used
to constrain the dominant backgrounds.
The four-top-quark production cross section is measured to be $24^{+7}_{-6}$~fb.
This corresponds to an observed (expected) significance with respect to the
background-only hypothesis of
4.3 (2.4) standard deviations and provides evidence for this process.
}
 
\author{The ATLAS Collaboration}
 
\AtlasRefCode{TOPQ-2018-05}

\PreprintIdNumber{CERN-EP-2020-111}
 
\AtlasJournalRef{Eur. Phys. J. C 80 (2020) 1085}
 
\AtlasDOI{10.1140/epjc/s10052-020-08509-3}

\hypersetup{pdftitle={ATLAS document},pdfauthor={The ATLAS Collaboration}}
 
\begin{document}
 
\maketitle
 
\tableofcontents
 
 
\section{Introduction}
 
Being the heaviest known elementary particle of the Standard Model (SM), the top quark has a large coupling to the SM Higgs boson
and is predicted to have large couplings to hypothetical new particles in many models
beyond the SM (BSM).
In  that respect, rare processes involving the top quark are particularly relevant to study.
Among these, the production of four top quarks (\tttt) is predicted by the SM but has not been observed yet.
The \tttt cross section is sensitive to the magnitude and CP properties of the Yukawa
coupling of the top quark to the Higgs boson since four top quarks can be produced via
an offshell SM Higgs boson~\cite{Cao:2017,Cao:2019ygh}.
Enhancements of the \tttt\ cross section (\stttt) are expected in many BSM scenarios, such as
gluino pair production in supersymmetry theories~\cite{Nilles:1983ge,Farrar:1978xj},
pair production of scalar gluons~\cite{Plehn:2008ae,Calvet:2012rk},
or the production of a heavy pseudoscalar or scalar boson in association with a top-quark pair (\ttbar) in Type II two-Higgs-doublet models (2HDM)~\cite{Dicus:1994bm,Craig:2015jba,Craig:2016ygr}.
Within an effective field theory framework~\cite{AguilarSaavedra:2018nen}, the BSM contribution to \tttt production can be parameterised by
non-renormalisable effective couplings and can be expressed for instance
via a \tttt contact interaction.
 
\begin{figure}[!htb]
\centering
\includegraphics[width=0.3\textwidth]{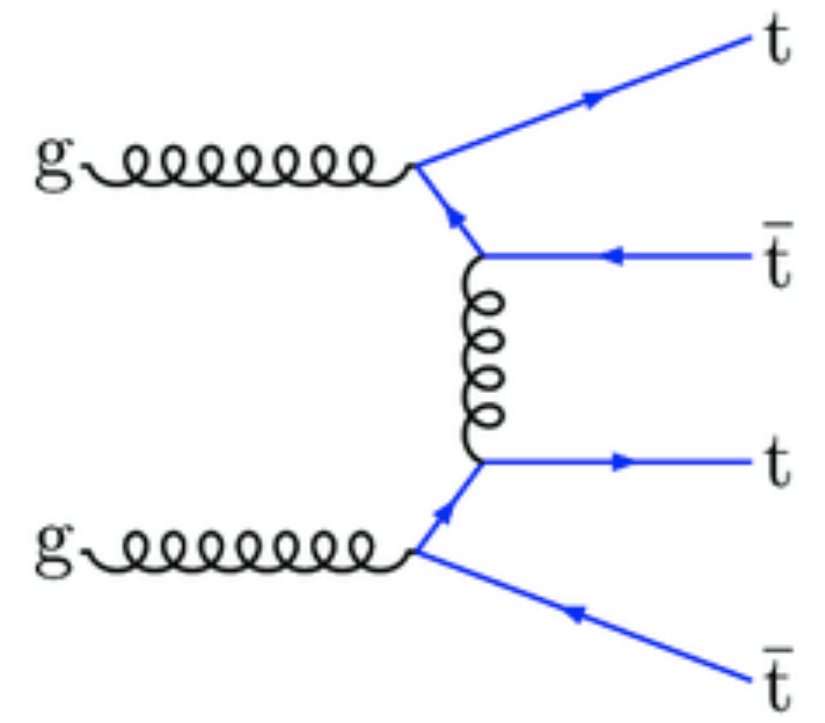}
\hspace{1 cm}
\includegraphics[width=0.3\textwidth]{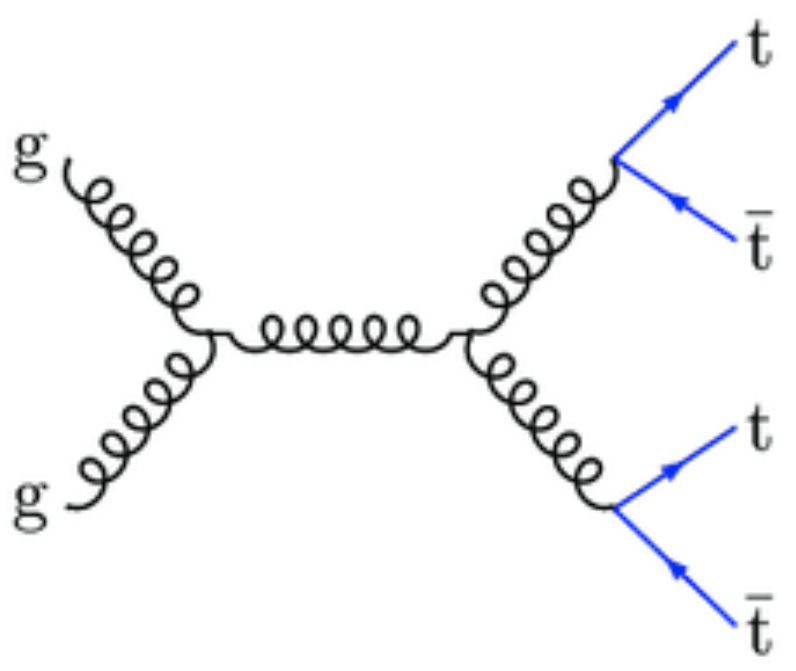}
\caption{Examples of Feynman diagrams for SM \tttt production at leading order in QCD.}
\label{fig:feynman}
\end{figure}

The cross section of the SM production of four top quarks from proton--proton ($pp$) collisions at
$\sqrt{s}=13$~\tev\ is predicted to be $\stttt = 12.0~\text{fb}$ with a relative scale uncertainty of $\pm20\%$
at next-to-leading order (NLO) in QCD including electroweak corrections~\cite{Frederix:2017wme}.
Examples of Feynman diagrams for \tttt QCD production in the SM are shown in Figure~\ref{fig:feynman}.
 
In the SM, the top quark is expected to decay into a $W$ boson and a $b$-quark with a
branching ratio of approximately 100\%.
Thus, the \tttt process will give rise to \WWWWbbbb events which
then produce different final states depending on the hadronic or leptonic decay mode of the $W$ bosons.
This paper considers events that contain exactly two isolated leptons with the same electric charge (2LSS)
or events with at least three isolated leptons\footnote{Throughout the paper, leptons refer to either electrons or muons, which can include those that come from a $\tau$-lepton decay.} (3L),
having branching fractions of 7\% and 5\%, respectively. Although
this channel, referred to as \SSML, has a small branching
fraction, it benefits from low levels of background.
The \tttt\ topology is characterised by high jet and $b$-jet multiplicities
and high overall energy, which can be quantified as a large value for the scalar
sum of the transverse momenta of objects in the event.
 
A number of SM processes can produce events with topologies similar to those of \tttt events and thus are backgrounds to \tttt production.
The dominant source is \ttbar production in association with other particles,
such as a Higgs boson (\ttH), $W$ boson (\ttW), or $Z$ boson (\ttZ).
Smaller contributions are expected from \ttWW, multi-boson production, single-top-quark as well as \ttt production.
Significant backgrounds also come from events where one of the leptons has a misassigned charge
and events that contain leptons arising from heavy-flavour decays, photon conversions or misidentified jets,
the latter three being collectively referred to as `\fake'.
The heavy-flavour decays are the dominant source for muons, while other sources mostly affect electrons.
The charge misassignment and \fake background comes mainly from \ttbar events.
 
In the analysis described in this paper, signal events are separated from background events using a multivariate discriminant.
A fit is then performed on the distribution of the multivariate discriminant
in the signal-enriched region.
Background-enriched regions are also added to the fit to determine the normalisations of the \ttW background and
of some sources of \fake background.
 
ATLAS and CMS previously searched for \tttt\  production in 13~\tev\ \(pp\) collisions.
The ATLAS search combined results in the \SSML channel with those in a channel comprising
single-lepton events and dilepton events with two opposite-sign charged  leptons
(called the \SLOS channel). This analysis used 36~\ifb\ of data and led to
an observed (expected) significance of 2.8 (1.0) standard
deviations~\cite{Aaboud:2018xpj,Aaboud:2018jsj}.
The CMS combination of the \SLOS and \SSML channels using 36~\ifb\ quotes an
observed (expected) significance of 1.4 (1.1) standard deviations~\cite{Sirunyan:2019nxl}.
The latest CMS search using 137~\ifb\ in the \SSML channel leads to an
observed (expected) significance for the \tttt\ signal of 2.6 (2.7) standard
deviations~\cite{Sirunyan:2019wxt}.
 
 
 
\section{The ATLAS detector}
 
\newcommand{\AtlasCoordFootnote}{
ATLAS uses a right-handed coordinate system with its origin at the nominal
interaction point (IP)
in the centre of the detector and the \(z\)-axis along the beam pipe.
The \(x\)-axis points from the IP to the centre of the LHC ring,
and the \(y\)-axis points upwards.
Cylindrical coordinates \((r,\phi)\) are used in the transverse plane,
\(\phi\) being the azimuthal angle around the \(z\)-axis.
The pseudorapidity is defined in terms of the polar angle \(\theta\) as \(\eta = -\ln \tan(\theta/2)\).
Angular distance is measured in units of \(\Delta R \equiv \sqrt{(\Delta\eta)^{2} + (\Delta\phi)^{2}}\).}
 
The ATLAS experiment~\cite{PERF-2007-01,PIX-2018-001,ATLAS-TDR-2010-19}
at the LHC is a multipurpose particle detector
with a forward--backward symmetric cylindrical geometry and a nearly \(4\pi\) coverage in
solid angle.\footnote{\AtlasCoordFootnote}
It consists of an inner tracking detector (ID) surrounded by a thin superconducting solenoid
providing a \SI{2}{\tesla} axial magnetic field, electromagnetic (EM) and hadron calorimeters, and a muon spectrometer.
The inner tracking detector covers the pseudorapidity range \(|\eta| < 2.5\).
It consists of silicon pixel, silicon microstrip, and transition radiation tracking detectors.
Lead/liquid-argon (LAr) sampling calorimeters provide electromagnetic (EM) energy measurements
with high granularity.
A steel/scintillator-tile hadron calorimeter covers the central pseudorapidity range (\(|\eta| < 1.7\)).
The endcap and forward regions are instrumented with LAr calorimeters
for both EM and hadronic energy measurements up to \(|\eta| = 4.9\).
The muon spectrometer (MS) surrounds the calorimeters and is based on
three large air-core toroidal superconducting magnets with eight coils each.
The field integral of the toroids ranges between \num{2.0} and \SI{6.0}{\tesla\metre}
across most of the detector.
The MS includes a system of precision tracking chambers and fast detectors for triggering.
A two-level trigger system is used to select events.
The first-level trigger is implemented in hardware and uses a subset of the detector information
to keep the accepted rate below \SI{100}{\kHz}.
This is followed by a software-based trigger that
reduces the accepted event rate to \SI{1}{\kHz}, on average,
depending on the data-taking conditions~\cite{TRIG-2016-01}.

 
 
\section{Object and event selection}
 
Data used in this analysis were collected by the ATLAS detector between 2015 and 2018 at $\sqrt{s} = \SI{13}{\TeV}$.
Only events for which all detector subsystems were operational are considered.
The data set corresponds to an integrated luminosity of
139~\ifb~\cite{ATLAS-CONF-2019-021,Avoni:2018iuv}.
 
Events were collected using single-lepton or dilepton triggers.
Single-lepton triggers select events with leptons satisfying either low transverse momentum (\pT) thresholds and
an isolation requirement, or a looser identification criterion and higher thresholds with no isolation requirement.
The lowest \pT\ thresholds used in the single-lepton triggers varied
from 20 to 26~\GeV\ depending on the lepton flavour and the data-taking
period~\cite{TRIG-2018-05,TRIG-2018-01}.
The \pT\ thresholds used in the dilepton triggers varied from 8 to 24~\GeV\
depending on the lepton flavour and the data-taking period.
Dilepton triggers are used to select events with leptons without requiring any isolation requirement; these are used to validate the \fake background estimation.
 
Events are required to have at
least one vertex reconstructed from at least two ID tracks with
transverse momenta of $\pT>\SI{0.4}{\GeV}$. The primary vertex for
each event is defined as the vertex with the highest sum of \pTsq\ over all associated ID tracks~\cite{ATL-PHYS-PUB-2015-026}.

Electron candidates are reconstructed from energy deposits in the EM
calorimeter associated with ID tracks~\cite{Aad:2019tso} and are required to have
a calorimeter energy cluster with pseudorapidity $|\eta_{\text{cluster}}|<2.47$,
excluding the transition region between the barrel and the endcap
calorimeters (${|\eta_{\text{cluster}}|\not\in[1.37,1.52]}$).
Muon candidates are reconstructed by combining tracks in
the ID with tracks in the MS~\cite{PERF-2015-10} and are required to have $|\eta|<2.5$.
Both the electron and muon candidates are required to have $\pT>$~28~\GeV.
The transverse impact parameter divided by its estimated
uncertainty, ${|\dzero|/\sigma(\dzero)}$, is required to be lower
than five (three) for electron (muon) candidates. The longitudinal impact
parameter must satisfy ${|\zzsth|<0.5}$~mm for both lepton flavours. Electrons are
required to meet the `Tight' likelihood-based identification criterion and to be
isolated using criteria based on the properties of the topological clusters
in the calorimeter and of the ID tracks around the reconstructed electron~\cite{Aad:2019tso}.
Muons are required to meet the `Medium' cut-based identification criterion,
which includes requirements on the number of hits in the ID and MS
as well as requiring compatibility between momentum measurements in the ID and MS.
Muons also have to satisfy the isolation requirement based on the properties of
ID tracks around the reconstructed muon.
 
To reduce the impact of charge misassignment background, an additional requirement is imposed on electrons
in the \SSee\ and \SSem\ channels. This requirement is based on the score of a boosted decision tree (BDT)
that uses the calorimeter cluster and track properties of the electron~\cite{Aad:2019tso} and is
trained on data enriched in $Z \to ee$ events to separate events with correct and incorrect electron charge assignments.
The chosen requirement on the BDT score removes approximately 90\% of electrons with
a wrong charge assignment while selecting 98\% of electrons with correctly measured charge.
 
Jets are reconstructed from topological clusters~\cite{PERF-2014-07} of energy
deposits in the calorimeters using the \antikt algorithm~\cite{Cacciari:2008gp,Fastjet}
with a radius parameter of \(R=0.4\) and are calibrated as described in
Ref.~\cite{PERF-2016-04}. Jets are required to have $\pT>\SI{25}{\GeV}$
and $|\eta|<2.5$.
To reduce the effect from additional \pp\ collisions in the same or a nearby bunch crossing,
collectively referred to as pile-up,
jets with $\pT<\SI{120}{\GeV}$ and $|\eta|<2.4$ are considered only when they satisfy a requirement
based on the output of a multivariate classifier called the jet-vertex-tagger (JVT)~\cite{PERF-2014-03}.
Events that contain at least one jet arising from non-collision sources or detector noise are rejected by a set of
quality criteria~\cite{ATLAS-CONF-2015-029}. The MV2c10 multivariate
algorithm~\cite{Aad:2019aic} is used to identify jets containing $b$-hadrons.
A jet is considered $b$-tagged if it passes the operating point corresponding to 77\% average
efficiency for $b$-quark jets in simulated \ttbar events with the corresponding rejection factors
against light-quark/gluon jets and $c$-quark jets of 110 and 4, respectively.
 
\newcommand{\rapidity}{
The rapidity is defined as $y=\frac{1}{2} \ln{\frac{E+p_z}{E-p_z}}$, where $E$ is the energy and $p_z$ is the component of the momentum along the beam pipe.
}
 
A sequential overlap removal procedure is applied to avoid the same calorimeter
energy deposit or the same track being reconstructed as two different objects. As a
first step, electrons sharing their track with a muon candidate are removed. Next,
the closest jet within $\Delta R_y = \sqrt{(\Delta y)^2+(\Delta \phi)^2} = 0.2$ of
an electron is removed.\footnote{\rapidity} Then, electrons within $\Delta R_y = 0.4$
of a remaining jet are removed since they likely arise from $b$- or $c$-decays.
After that, jets with fewer than three associated tracks that are within $\Delta R_y = 0.2$ of a muon are removed.
Finally, muons are removed if their tracks are within $\Delta R_y = 0.4+ \SI{10}{\GeV}/{\pT}_{\mu}$ of any remaining jets
as they also likely arise from $b$- or $c$-decays.
 
The missing transverse momentum in the event, whose magnitude is denoted in the
following by \met, is defined as the negative vector sum of the \pT\ of the reconstructed
and calibrated objects in the event~\cite{Aaboud:2018tkc}. This sum includes the momenta of the ID tracks
that are matched to the primary vertex and are not associated with any other objects.
 
The events are required to have one same-sign lepton pair or at least three leptons without charge requirement.
Each event must have at least one reconstructed lepton that matches a lepton that fired the trigger.
Events with two same-sign electrons are required to have the invariant mass
$m_{ee}>15~\GeV$ and $|m_{ee}-91~\GeV|>10~\GeV$ to reduce the charge
misassignment background coming from low-mass resonances and $Z$-boson decay.
In events with at least three leptons, all the opposite-sign same-flavour lepton pairs are
required to satisfy $|\mll-91~\GeV|>10~\GeV$ to reduce the contamination from $Z$-boson decay.
 
Events arising from \tttt production are selected by exploiting the
high multiplicities of light-flavour jets and $b$-tagged jets
as well as the large overall event activity.
This last property is probed by the scalar sum of the transverse momentum
of the isolated leptons and jets in the event, denoted by \HT.
The inclusive signal region (SR) is defined by requiring at least six jets,
at least two $b$-tagged jets, and \HT above $500\,\GeV$.

 
\section{Monte Carlo samples}
\label{sec:mcsamples}
 
Production of \tttt events is modelled according to the SM expectation.
The nominal sample used to model the \tttt signal was generated using the \MGMCatNLOV{v2.6.2} \cite{Alwall:2014hca}
generator, which provides matrix elements (ME) at NLO in the strong coupling
constant \alphas, with the \nnpdfone~\cite{Ball:2014uwa}
PDF set.
The functional form of the renormalisation and factorisation scales was set
to $\muR = \muF = m_{\texttt{T}}/4$,
where $m_{\texttt{T}}$ is defined as the scalar sum of the transverse masses
$\sqrt{p_{\texttt{T}}^{2} + m^{2}}$ of the particles generated from the matrix
element calculation, following Ref.~\cite{Frederix:2017wme}.
The parton shower, fragmentation, and underlying event were simulated using
\PYTHIAV{8.230}~\cite{Sjostrand:2014zea} with the A14 set of tuned parameters
(tune)~\cite{ATL-PHYS-PUB-2014-021} and the \nnpdftwo PDF set.
The top-quark mass \mtop in this sample and in all
other simulated samples is set to 172.5~\gev.
An alternative \tttt sample generated with the same \MGMCatNLO set-up but interfaced
to \HerwigV{7.04}~\cite{Bahr:2008pv,Bellm:2015jjp} with the H7UE tune~\cite{Bellm:2015jjp}
and the \texttt{MMHT2014LO} PDF set~\cite{Harland-Lang:2014zoa} is used to evaluate
uncertainties due to the choice of parton shower and hadronisation model.
In order to mitigate the effect
of the large fraction of negative weights present in the nominal sample that would be
detrimental for
training of the multivariate discriminant used to separate signal from background
(see Section~\ref{sec:strategy}),
an additional sample with settings similar to the nominal ones
was generated using leading-order (LO) matrix elements.
Good agreement between the distributions of the kinematic variables used by
the multivariate discriminant simulated at LO and NLO was observed.

The $t\bar{t}W$ simulated events were generated using the \SHERPAV{2.2.1}~\cite{Bothmann:2019yzt} generator with
the \nnpdfnlo PDF set and the tune provided by the \SHERPA authors.
The ME was calculated for up to one additional
parton at NLO QCD and up to two partons using the five-flavour scheme, including $c$- and $b$-quarks, at LO QCD using the Comix~\cite{Bothmann:2019yzt} and
OpenLoops~\cite{Cascioli:2011va,Denner:2016kdg} libraries, and was merged with the \SHERPA parton
shower~\cite{Schumann:2007mg} using the MEPS@NLO
prescription~\cite{Hoeche:2011fd,Hoeche:2012yf,Catani:2001cc,Hoeche:2009rj} and a merging scale of 30~\GeV.
The renormalisation and factorisation scales were set to
$\muR = \muF = m_{\texttt{T}}/2$.
The simulated $t\bar{t}W$ sample is normalised to the cross section of 601~fb computed at NLO in QCD with the
leading NLO electroweak corrections~\cite{Campbell:2012dh,Frixione:2015zaa,deFlorian:2016spz}.
An alternative $t\bar{t}W$ sample was generated at NLO in QCD with no additional partons using the
\MGMCatNLOV{v2.3.3} generator with the same PDF as the nominal sample.
The events were interfaced to \PYTHIAV{8.210} using the A14 tune
and the \nnpdftwo PDF set.
 
The production of \ttdy events was modelled at NLO in QCD using the \MGMCatNLO~v2.3.3 generator
in the five-flavour scheme with the \nnpdfnlo PDF set interfaced to \PYTHIAV{8.230} using
the A14 tune and the \nnpdftwo PDF set.
It is normalised to
the inclusive $\ttll$ cross section of 880~fb computed at
NLO in QCD~\cite{Campbell:2012dh,Frixione:2015zaa,deFlorian:2016spz},
including off-shell $Z$ and $\gamma^*$ contributions with $m(\ell^+\ell^-) > 5~\gev$.
An alternative sample was generated with NLO matrix elements
using \SHERPA~2.2.1 and the same PDF as the nominal sample.
 
The production of \ttbar\ and single-top-quark events was modelled using
the \POWHEGBOX~\cite{Frixione:2007nw,Nason:2004rx,Frixione:2007vw,Alioli:2010xd}~v2 generator
at NLO in QCD with the \nnpdfnlo PDF set. In the \ttbar\ sample
the \hdamp\ parameter\footnote{The \hdamp\ parameter controls the transverse momentum \pt\ of the first
additional emission beyond the leading-order Feynman diagram in the parton shower and therefore
regulates the high-\pt\ emission against which the \ttbar\ system recoils.} was set
to 1.5~\mtop~\cite{ATL-PHYS-PUB-2016-020}.
The overlap between the \ttbar\ and the $tW$ final
states was removed using the diagram removal technique~\cite{Frixione:2008yi}.
The \ttbar\ and single-top-quark simulated samples are normalised to the cross sections
calculated at next-to-next-to-leading order (NNLO) in QCD including the
resummation of next-to-next-to-leading logarithmic (NNLL) soft-gluon terms
~\cite{Czakon:2011xx,Kidonakis:2010ux,Kidonakis:2010tc,Kidonakis:2011wy}.
 
The production of $t\bar{t}H$ events was modelled by
the \POWHEGBOXV{v2} generator at NLO in QCD using the five-flavour scheme with the
\nnpdfnlo PDF set, with the \hdamp parameter set to $1.5\times(2\mtop+m_H)/2$
where the Higgs boson mass is $m_H=125$~\gev.
The simulated sample is normalised to the cross section
computed at NLO in QCD with the leading NLO electroweak
corrections~\cite{Campbell:2012dh,Frixione:2015zaa,deFlorian:2016spz}.
An alternative sample generated using the \MGMCatNLOV{v2.3.3} generator with
the same settings is used to evaluate the uncertainty in $t\bar{t}H$ modelling due to the
generator choice.
 
The \tWZ events were generated at NLO in QCD using the \MGMCatNLOV{v2.3.3} generator with the
\nnpdfnlo PDF set.
The other rare top-quark processes, namely the production
of \tZ, \ttWW \ttZZ, \ttWZ, \ttHH and \ttWH, and \ttt, were
modelled using the \MGMCatNLO generator at LO in QCD. The \ttbar,
single-top-quark, $t\bar{t}H$, \tZ, \tWZ, \ttWW, \ttZZ, \ttWZ, \ttHH, \ttWH
and \ttt generated samples were all interfaced with \PYTHIAV{8.230} using
the A14 tune and the \nnpdftwo PDF set.
Rare top-quark background contributions are normalised using their NLO QCD
theoretical cross sections. The \ttt production is normalised to the
cross section of 1.64~fb calculated at LO in QCD with \nnpdftwo.
 
The $WH$ and $ZH$ processes were generated using
the \PYTHIAV{8.230} generator with the A14 tune and
\nnpdftwo PDF set and normalised to their theoretical cross sections
calculated at NNLO in QCD and NLO electroweak accuracies.
 
Samples of diboson ($\VV$) and triboson (\VVV) production
were simulated with the \SHERPAV{2.2.1} generator,
and samples of $Z$+jets and $W$+jets production were simulated with the
\SHERPAV{2.2.2} generator,
both with the \nnpdfnnlo PDF set.
The $\VV$ and \VVV samples are normalised to the theoretical
cross sections calculated at NLO QCD, and $Z$+jets and $W$+jets backgrounds
are normalised to the NNLO cross sections~\cite{Aad:2016naf}.
 
The effects of pile-up were modelled by overlaying minimum-bias events,
simulated using the soft QCD processes of \PYTHIAV{8.186}
with the A3 tune~\cite{ATL-PHYS-PUB-2016-017}, on events from hard processes.
For all samples of simulated events, except those generated using \SHERPA, the EvtGen v1.2.0
program~\cite{Lange:2001uf} was used to describe the decays of bottom and charm hadrons.
 
The nominal signal and background samples were processed through the
simulation~\cite{Aad:2010ah} of the ATLAS detector geometry and response
using \textsc{Geant4}~\cite{Agostinelli:2002hh}, and then reconstructed using the same software as is used for the collider data.
Some of the alternative samples used to evaluate systematic uncertainties were
instead processed through a fast detector simulation making use of parameterised showers
in the calorimeters~\cite{ATL-PHYS-PUB-2010-013}.
Corrections were applied to the simulated events so that the physics objects'
selection
efficiencies, energy scales and energy resolutions match those determined from data
control samples.

 
\section{Background estimation}
\label{sec:backSSML}
 
Backgrounds in the \SSML channel can be categorised as irreducible
and reducible.
Irreducible backgrounds are those for which all selected leptons are from $W$- or $Z$-boson decays
or from leptonic $\tau$-lepton decays.
The main irreducible
backgrounds originate from the \ttW, \ttZ and \ttH processes, mainly when
additional jets are $b$-jets. The smaller backgrounds include diboson or triboson production, $VH$ production
in association with jets, and rare processes (\ttWW, \tWZ, \tZq, \ttt). The irreducible
background is evaluated using MC simulation normalised to the SM cross sections, except \ttW for which the normalisation is corrected using
data in a dedicated control region.
The different treatment for the \ttW background is motivated by theoretical studies~\cite{Frederix:2020jzp}
showing that electroweak corrections not included in the simulation have a significant effect as well as by the
large \ttW background normalisation factor found in recent measurements in similar phase space~\cite{ATLAS-CONF-2019-045}.
 
The reducible backgrounds originate mainly from \ttjets and $tW$+jets
production and have prompt leptons with misassigned charge (\Qmis) or \fake leptons.
This \fake background, together with \ttW, is evaluated using the template method (cf.\ Section~\ref{sec:fakes}).
The charge misassignment background is defined for the 2LSS channels only.
It arises mainly from \ttjets events with an opposite-charge lepton pair in which the charge
of one electron is mismeasured either due to bremsstrahlung photon emission followed by its conversion
($e^{\pm} \to e^{\pm} \gamma \to e^{\pm} e^+ e^-$) or due to mismeasured track curvature.
This background is evaluated using a data-driven method (cf.\ Section~\ref{sec:qmisid}).
The charge misassignment rate is negligible for muons due to the low
probability of bremsstrahlung and the large lever arm of the muon spectrometer.
 
The estimated yield from each source of background is given in Section~\ref{sec:results}.

 
\subsection{Fake/non-prompt lepton background and \ttW production}
\label{sec:fakes}
 
The template method used to estimate the \fake background relies on the
simulation to model the kinematic distributions of background processes arising from
fake and non-prompt leptons and on control regions to determine their
normalisations. These control regions are included in the fit together
with the signal region, and the normalisation factors are determined
simultaneously with the \tttt signal.
 
The following main contributions of the \fake background are distinguished:
\begin{itemize}
\item events with one non-prompt electron (muon) from heavy-flavour decay, HF~$e$ (HF~$\mu$),
\item events with one non-prompt electron originating from photon conversion
taking place in the detector material (Mat.\ Conv.),
\item events with a virtual photon ($\gamma^*$) leading to an \ee pair (Low $m_{\gamma^*}$).
\end{itemize}
The minor components of the \fake background arising from events with
a lepton originating from light-meson decay (LF) or with a jet misidentified
as a lepton (other fakes) are determined from MC simulation.
 
Several control regions, non-overlapping with the signal region, are defined to determine the normalisation
of various components of the \fake background from data.
Each region is required to have a dominant component or a variable with good discriminating power
between different components.
Since events arising from \ttW production represent a large contribution
in all control and signal regions, the normalisation of that process is also determined
using a dedicated control region.
In total, four control regions with their corresponding discriminating
variables are used in the analysis. They are summarised in Table~\ref{tab:CRSR_SSML}
and are defined below:
 
\begin{itemize}
 
\item `\CO' is enriched in background events arising from both material photon conversion and processes
with a virtual photon leading to an $e^{+}e^{-}$ pair.
For each electron in the selected $e^\pm e^\pm$ or $e^\pm \mu^\pm$ events,
the invariant mass of the system formed by the track associated with
the electron and the closest track at the conversion (primary) vertex
$m_{ee}^\mathrm{CV}$ ($m_{ee}^\mathrm{PV}$) is computed.
The conversion vertex is defined as the point where the track from the electron and its closest track in $\Delta R$ have the same $\phi$.
The control region is then obtained
by selecting events
with at least four or five jets, at least one identified b-jet,
with low $m_{ee}^\mathrm{CV}$ and using the $m_{ee}^\mathrm{PV}$ distribution
in the fit to separate the material conversion and the $\gamma^*$ components from each other.
Virtual photons lead to a lepton pair originating from the primary vertex, having
a low $m_{ee}^\mathrm{PV} \sim m_{\gamma^*}$ and a low conversion radius.
Material conversions happen further away from the primary vertex with a larger conversion radius, and the track extrapolation induces
a larger apparent invariant mass. According to the MC simulation, the background
arising from both $\gamma^*$ and material conversions
accounts for around $40\%$ of the total event yield in this control region.
 
\item `\Le' (`\Lmu') is enriched in background events with an electron (muon) from heavy-flavour decay.
This region is defined by selecting events with three leptons, namely $eee$ and $ee\mu$ ($\mu\mu\mu$ and $\mu\mu e$)
for \Le (\Lmu), and exactly one identified $b$-jet. This selection targets \ttbar\ dileptonic decays
with an extra non-prompt lepton in events with low \Ht.
The number of events in the region is used in the maximum-likelihood fit.
According to the MC simulation, the background with an electron (muon)
coming from heavy-flavour decay accounts for around  $40\%$ ($50\%$) of the total event yield in
the \Le (\Lmu).
 
\item `\CRttW' is enriched in \ttW events.
This region is obtained by selecting $e\mu$ and $\mu\mu$ events with at least four jets and two $b$-jets
which are neither in other CRs nor in the SR.
Events containing electrons with $|\eta|>1.5$ and $ee$ final states are not considered, in order to reduce
the contamination arising from charge misassignment background.
The sum of the lepton \pt provides discrimination from other processes and is used in the maximum-likelihood fit.
According to the MC simulation, the \ttW background accounts for
around  $33\%$ of the total event yield in this control region.
 
\end{itemize}
 
 
\begin{table}[htb]
\renewcommand*{\arraystretch}{1.3}
\begin{small}
\begin{center}
\caption{Summary of the signal and control regions used in the template fit.
The variable $m_{ee}^\mathrm{CV}$ ($m_{ee}^\mathrm{PV}$) is defined as
the invariant mass of the system formed by the track associated with the
electron and the closest track at the conversion (primary) vertex.
$N_j$ ($N_b$) indicates the jet ($b$-tagged jet) multiplicity in the event.
$\Ht$ is defined as the scalar sum of the transverse momenta of the isolated leptons and jets.}
\begin{tabular}{|c|c|c|c|c|c|}
\hline
Region    & Channel           & $N_j$               & $N_b$   & Other requirements      & Fitted variable \\ \hline \hline
SR        & \SSML             & $\ge 6$             & $\ge 2$ & $\Ht > 500$             & BDT      \\ \hline \hline
\CO    & $e^\pm e^\pm || e^\pm \mu^\pm$    & $4 \leq N_j < 6$    & $\ge 1$ & $m_{ee}^\mathrm{CV} \in [0,0.1~\GeV]$ & $m_{ee}^\mathrm{PV}$ \\
&                   &                     &         & $200 < \Ht < 500~\GeV$       &          \\ \hline
\Le    & $eee$ $||$ $ee\mu$      &           -          & $=1$    & $100<\Ht<250~\GeV$           & counting  \\ \hline
\Lmu   & $e\mu\mu$ $||$ $\mu\mu\mu$      &     -                & $=1$    & $100<\Ht<250~\GeV$           & counting  \\ \hline
\CRttW & $e^\pm \mu^\pm || \mu^\pm \mu^\pm$    & $\ge 4$        & $\ge 2$ & $m_{ee}^\mathrm{CV} \not\in [0,0.1~\GeV]$, $|\eta(e)|<1.5$ & $\Sigma \ptl $ \\
&                   &                     &         & for $N_b = 2$, $\Ht < 500~\GeV$ or $N_j < 6$&      \\
&                   &                     &         & for $N_b \ge 3$, $\Ht < 500~\GeV$ &      \\ \hline
\end{tabular}
\label{tab:CRSR_SSML}
\end{center}
\end{small}
\end{table}
 


 
\subsection{Charge misassignment background}
\label{sec:qmisid}
 
The probability for an electron to have its charge incorrectly assigned is
measured using a data sample of $Z \to ee$ events requiring the invariant mass of
the electron pair to be within 10~\gev\ of the $Z$-boson mass and
without any requirement on the charge of the two electron tracks.
The background contamination is subtracted using a sideband
method~\cite{Aaboud:2018xpj}.
The charge misassignment rate is parameterised as a function of
electron \pT and $|\eta|$, except for the
conversion control region defined in Section~\ref{sec:fakes}, where
it is also parameterised as a function of the invariant mass of
the electron track and its closest track assuming that both tracks
originate from the primary vertex.
The charge misassignment rate varies from 0.002\% to 4\%
depending on the electron \pT and $|\eta|$.
 
The expected number of events arising from charge misassignment background is determined by
applying the measured charge misassignment rate to data events satisfying the
requirements of the kinematic selection of the 2LSS channel, except that the two leptons are required to be
of opposite charge. In this sample, each event is weighted according to
the value of the charge misassignment rate of each electron in the event.


 
 
\section{Signal discrimination}
\label{sec:strategy}

The background composition of the SR is largely dominated by the production of
top-quark pairs in association with additional jets and/or bosons.
To separate signal from background events,
a multivariate discriminant is built in the SR by combining several input
observables into a boosted decision tree (BDT).
This set of variables and the BDT hyperparameters are optimised
to maximise the integral under the receiver operating characteristic (ROC)
curve of the BDT. In total, 12 observables are selected, based on their
discrimination power and the requirement of good modelling.
Among them, the pseudo-continuous $b$-tagging discriminant
score~\cite{Aad:2019aic} summed
over all the jets in the event is the best discriminating variable
due to the four $b$-jets being produced mainly in signal events.
The pseudo-continuous $b$-tagging discriminant score is an integer from 1 to 5
assigned to a jet, based on the operating point of the $b$-tagging algorithm
it passes, with a value of five corresponding to the most $b$-like jet.
For the signal region selection the minimum score is 10 for an event with exactly two $b$-tagged jets out of exactly six jets.
The minimum distance $\Delta R = \sqrt{(\Delta \eta)^2+(\Delta \phi)^2}$
between two leptons among all possible pairs is the second best discriminating variable since it
provides good discrimination for events with at least three leptons.
The distributions for those two variables are shown in
Figure~\ref{fig:prefitinputvar}.
The other input variables are
the leading lepton $\pT$,
\met,
the \pT of the leading and second-leading jets,
the \pT of the sixth jet,
the \pT of the leading $b$-jet,
the scalar sum of transverse momenta over all leptons and all jets excluding the leading $\pT$ jet,
the sum of distances $\Delta R$ between two leptons for all possible pairs,
the maximum distance $\Delta R$ between a $b$-jet and a lepton among all possible pairs,
and the minimum distance $\Delta R$ between a jet and a $b$-jet among all possible pairs.
Taking into account all uncertainties, no significant discrepancy between
data and predicted background was found for these variables in the various CRs.
 
\begin{figure}[!htb]
\centering
\includegraphics[width=0.45\textwidth]{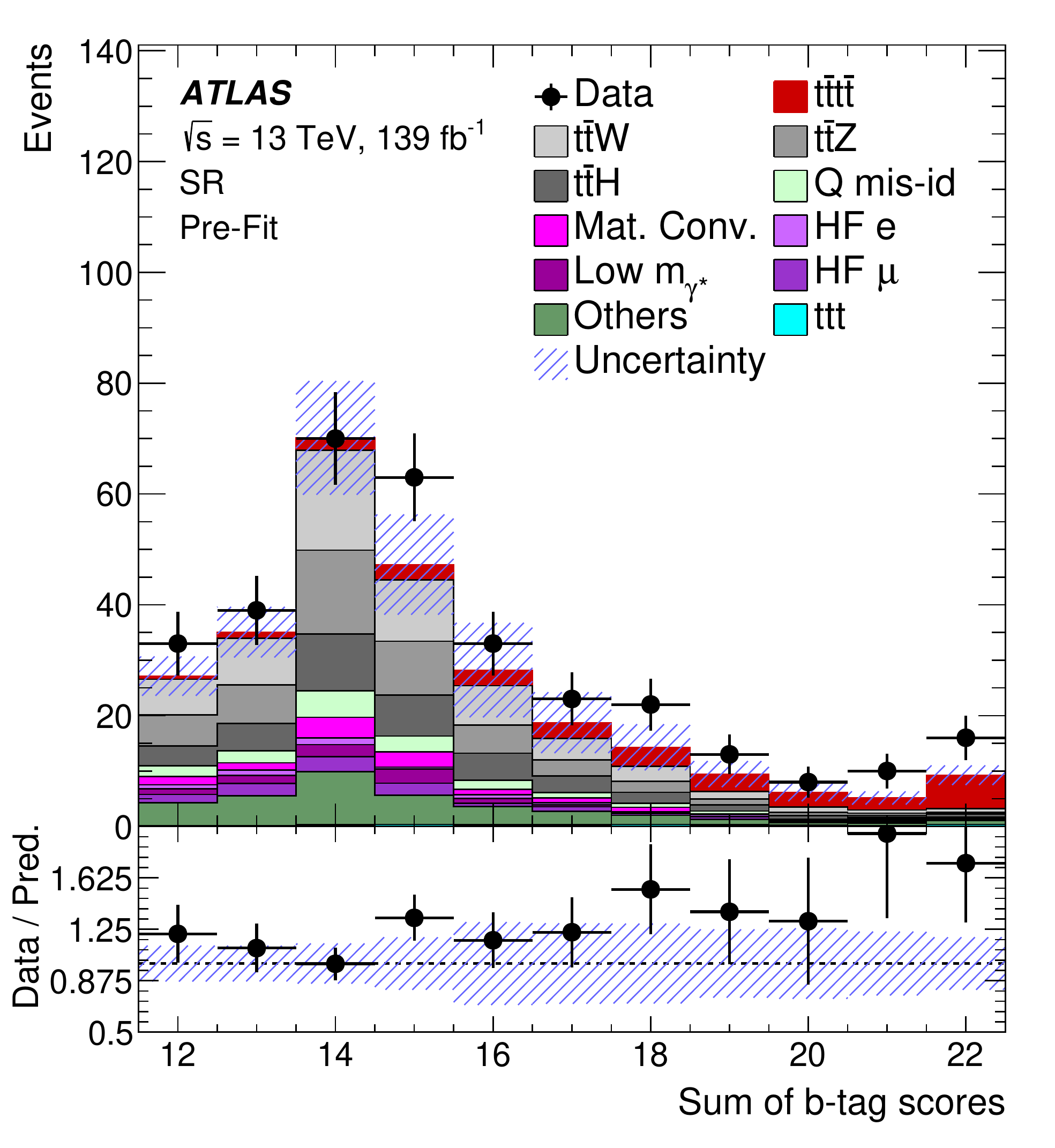}
\includegraphics[width=0.45\textwidth]{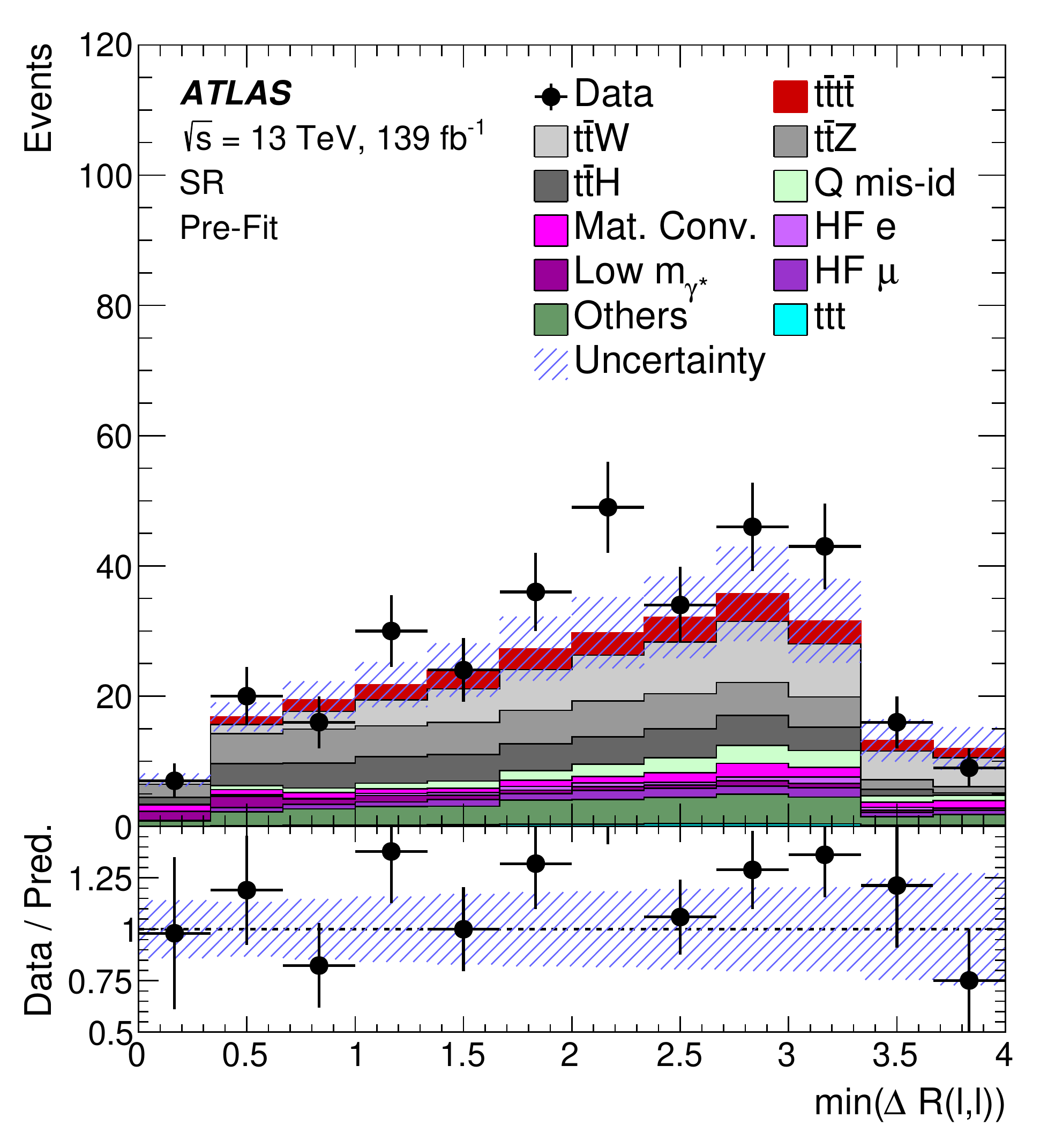}
\caption{Pre-fit comparison between data and prediction in the signal region
for two of the input variables used to train the multivariate discriminant:
the pseudo-continuous $b$-tagging discriminant score summed over all the jets
in the event (left) and
the minimum distance between two leptons among all possible pairs (right).
The band includes the total uncertainty of the pre-fit computation.
The ratio of the data to the total pre-fit expectation is shown in the lower panel.
The first and last bins contain underflow and overflow events, respectively.
See Section~\ref{sec:backSSML} for the definitions of the different background categories.
}
\label{fig:prefitinputvar}
\end{figure}
 
The BDT training is performed inclusively, both in lepton flavour and
lepton multiplicity for events passing the SR requirements.
The LO \tttt simulated signal sample is used in the training.
The background sample corresponds to the total expected background, as predicted by
the simulation.

 
 
\section{Systematic uncertainties}
\label{sec:systematics}
Various sources of systematic uncertainty
impact the estimated signal and background
rates, the migration of events between regions, and the shape of the fitted
discriminant distributions. They can be classified into the experimental uncertainties and
modelling uncertainties of the \tttt signal and of backgrounds, as described below.
The impact of each source of systematic uncertainty on the final result can be found in Section~\ref{sec:results}.
 
\subsection{Experimental uncertainties}
 
The uncertainty in the combined 2015--2018 integrated luminosity
is 1.7\%~\cite{ATLAS-CONF-2019-021}, obtained using the LUCID-2 detector~\cite{Avoni:2018iuv} for the primary luminosity measurements.
To account for the difference between the pile-up distributions in data and MC simulations, an uncertainty related to the scale factors
used to adjust the MC pile-up to the data pile-up profile is applied.
 
For electrons and muons, the reconstruction, identification, isolation and
trigger performances as well as the lepton momentum scale and resolution
differ between data and MC simulation.
To correct for these differences, scale factors for each are applied.
These scale factors were estimated using the tag-and-probe
method~\cite{Aad:2019tso,PERF-2015-10}, which is performed using the leptonic
decays of $Z$ and $W$ bosons and of $J/\psi$ mesons.
The associated systematic uncertainties are then propagated to the final distributions used in this analysis.

To determine the jet energy scale (JES) and its associated uncertainty, information from
test-beam data, LHC collision data, and simulation was used, as described
in Ref.~\cite{ATL-PHYS-PUB-2015-015}. The JES uncertainty
is decomposed into a set of 30 uncorrelated components, 29 of which are used per event
depending on the type of simulation.
The jet energy resolution (JER) is measured separately for data and MC simulation
using \emph{in situ} techniques, similar to those in Ref.~\cite{Aad:2011he}. Its uncertainty
is represented by nine components accounting for jet-$\pt$ and $\eta$-dependent differences between
simulation and data, eight of which are used per event depending on the type
of simulation.
The systematic uncertainty associated with the JVT is obtained by varying the scale factor used to correct the JVT efficiency in simulation up and down within its uncertainties~\cite{PERF-2014-03}.
 
The $b$-tagging efficiencies and mistagging rate are measured in data using the same methods as are described in Refs.~\cite{Aad:2019aic,ATLAS-CONF-2018-001,ATLAS-CONF-2018-006},
with the systematic uncertainties due to $b$-tagging efficiency and the mistagging rates calculated separately.
The impact of the uncertainties on the $b$-tagging calibration is evaluated separately for $b$-jets, $c$-jets and light-flavour jets in the MC samples.
 
The \met uncertainty due to a possible miscalibration of its soft-track component is derived
from data--MC comparisons of the \pT balance between the hard and soft \met components~\cite{Aaboud:2018tkc}.
 
 
\subsection{Signal modelling uncertainties}
\label{sec:SystSignal}
Several sources of modelling uncertainty are considered for
the \tttt signal. The uncertainty due to
missing higher-order QCD corrections is determined by
varying the renormalisation and the factorisation scales simultaneously by
factors of 2.0 and 0.5 relative to the central value.
The uncertainty related to the choice of parton shower and hadronisation
model is estimated by comparing the nominal prediction with that obtained
using an alternative sample generated with \MGMCatNLO interfaced to
\HerwigV{7}.
The effect of the PDF uncertainty on the signal MC prediction is calculated as
the RMS of the signals from the 100 replicas of the $\texttt{NNPDF30\_nlo\_as\_0118}$
PDF set following the PDF4LHC prescription~\cite{Botje:2011sn}.
Shape and normalisation variations due to the PDF uncertainty
are found to be negligible.
 
\subsection{Modelling uncertainties in irreducible background}
 
Modelling uncertainties for the \ttW, \ttZ and \ttH processes are evaluated in a similar
way and include the uncertainty due to
missing higher-order QCD corrections determined by
varying the renormalisation and the factorisation scales simultaneously by
factors of 2.0 and 0.5 relative to the central value, and a comparison with alternative generators.
 
For \ttZ, the nominal MC prediction is compared with an NLO \sherpa sample,
while for \ttH the nominal simulation is compared with an NLO \MGMCatNLO sample, both described in
Section~\ref{sec:mcsamples}.
A 1\% uncertainty from the PDF is assigned to both the $t\bar{t}Z$ and $t\bar{t}H$ processes
following the same procedure as described in Section~\ref{sec:SystSignal}.
For \ttW, uncertainties associated with the modelling of
additional QCD radiation, with the choice of the ME generator and parton shower, are estimated
by comparing the nominal prediction with that of an alternative sample that was generated at NLO
with no additional partons using the \MGMCatNLO generator
with the same scale choice and PDF set as for the nominal sample (cf.\ Section~\ref{sec:mcsamples}).
 
An uncertainty of $15\%$ ($20\%$) is applied to the $t\bar{t}Z$ ($t\bar{t}H$) total cross section~\cite{deFlorian:2016spz,Aad:2019mbh}.
Since the \ttW normalisation is determined from the fit to data, no cross-section uncertainty is
applied to this process. An additional $125\%$ (300\%) uncertainty is added for $t\bar{t}W$ production with
seven (eight or more) jets. These values correspond to the difference between data and
prediction in a \ttW validation region (cf.\ Section~\ref{sec:results})
where a data excess is observed for high jet multiplicities.
Since the jet multiplicity distribution in a \ttZ validation region shows good agreement
between data and prediction, such uncertainty is not considered for \ttZ or for \ttH production due to similarity of their simulation.
 
The \ttW, \ttZ and \ttH background processes enter the \tttt signal region if they have additional heavy-flavour jets.
Such processes are difficult to model with the MC simulation.
To account for this, an uncertainty of 50\% is assigned to the
events with three generator-level (`true') $b$-jets and a separate 50\%
uncertainty to the events
with four or more true $b$-jets.
These estimates are based on the measurement of
\ttbar production with additional heavy-flavour jets~\cite{TOPQ-2017-12}
and on comparisons between data and prediction in $\ttbar \gamma$
events with three and four $b$-tagged jets.
They are treated as uncorrelated between the three backgrounds due to the different MC setups used
to simulate the \ttW, \ttZ and \ttH backgrounds.
 
The \ttt events have similar kinematics to the \tttt signal, although the rate is expected
to be much smaller. However, it is currently unexplored experimentally.
Thus a large ad hoc uncertainty of $100\%$ is assigned to its cross section and an additional
$50\%$ uncertainty is applied to \ttt events with four true $b$-jets.
 
The uncertainty in the $tZ$ and $tWZ$ single-top-quark cross sections is set to
30\%~\cite{Aaboud:2017ylb,Demartin:2016axk} and that for the $t\bar{t}WW$, \ttZZ, \ttWZ, \ttHH
and \ttWH cross sections to 50\%~\cite{Aaboud:2018xpj}.
The uncertainty in diboson production is set to 40\%, based on studies of the $WZ+b$ process.
For each of the other small background processes a large ad hoc cross-section uncertainty of
$50\%$ is applied. For all small backgrounds except \ttt an additional 50\%
uncertainty is assigned to the events with three true $b$-jets and separately
a 50\% uncertainty for events with four or more true $b$-jets.
 
 
\subsection{Modelling uncertainties in reducible background}
 
Uncertainties in the charge misassignment background arise from the following contributions:
the statistical uncertainty
of the fit to data used to determine the rates; the rate variation
due to variation of the dielectron invariant mass requirement; and the rate variation
due to a difference between the observed and the predicted misidentification
rates when the method is applied to MC simulated events.
This uncertainty is determined separately for the material conversion control region,
for the \ttW, and for all other control regions, and it is treated as
correlated between the regions.
 
Since the overall normalisations of the material conversion and the virtual photon backgrounds are free parameters
in the fit, their uncertainty comes only from the shape of the
distributions used in the template fit (cf.\ Section~\ref{sec:fakes}).
For each of these sources, the uncertainty is obtained by comparing
data with the $\POWPYTHIA8$ simulation of $Z(\to\mu\mu)+\gamma$ and $Z(\to\mu\mu)+$jets production
in a region enriched in $Z(\to\mu\mu)+\gamma$ events.
An uncertainty of 25\% is applied to the material conversion and to the virtual photon background events fulfilling
$m_{ee}^\mathrm{CV}>0.1~\GeV$ in all control and signal regions
to cover the extrapolation from the `\CO' region with
$0<m_{ee}^\mathrm{CV}<0.1~\GeV$ to the regions with events with larger
$m_{ee}^\mathrm{CV}$.

The uncertainty in the shape of the distributions of the heavy-flavour non-prompt lepton background
is estimated by comparing data with the background
prediction, normalised to data, for a loose lepton selection
with the isolation requirements dropped and the identification
criteria relaxed. The shape uncertainty is derived for each region included in the
fit, but
these variations are treated as correlated between regions
since the physics origin of the uncertainty is
common to all of them.
This systematic uncertainty is derived separately for
electrons and muons.
 
A normalisation uncertainty of $100\%$ is assigned to the background arising from light-flavour non-prompt leptons.
This uncertainty was found to cover any difference between data and prediction in loose lepton regions~\cite{ATLAS-CONF-2019-045}.
An ad hoc uncertainty of $30\%$ is applied to the normalisation of the
background arising from the other minor sources of non-prompt leptons
from \ttbar production. No uncertainty in the shape of the distributions
of these backgrounds is considered since their contribution is very small.
 
Since the main source of reducible background is \ttjets production,
the systematic uncertainty in the modelling of its heavy-flavour content
can affect the shape of the template distributions used in the fit. To account
for this effect an uncertainty of 30\%, based on the measurement of
\ttbar production with additional heavy-flavour jets~\cite{TOPQ-2017-12},
is assigned to the events with three true $b$-jets and a separate 30\% uncertainty
to events with four or more true $b$-jets.
A small contribution to the reducible background from \Vjets production is
determined
from simulation and a normalisation uncertainty of $30\%$ is assigned to this background.
 
 
 
 
\section{Results}
\label{sec:results}
 
The \tttt production cross section and the normalisation factors of the
backgrounds are determined via a binned likelihood fit
to the BDT score distribution in the signal region and to the yields, or
to the discriminating
variable distributions, in the four control regions as listed in Table~\ref{tab:CRSR_SSML}.
The systematic uncertainties in both the signal and background predictions are
included as nuisance parameters in the likelihood function.
The maximum-likelihood fit is performed using the \texttt{RooFit}
package~\cite{RooFit} based on a likelihood function
built on the Poisson probability that the observed data are compatible with
the model prediction. The value of each nuisance parameter is
constrained by a penalty factor present in the likelihood function, while all
normalisation factors are unconstrained.
 
The fit determines the best value of the signal strength $\mu$, defined as a ratio of
the \tttt cross section to the SM expectation, its uncertainty, and five
normalisation factors: \NFHFe (\NFHFm) for the non-prompt electron (muon) background
from heavy-flavour decays, \NFCO for the background from detector material conversions,
\NFGst for the contribution of low-mass electron pairs, and \NFttW for the \ttW contribution.
For each free parameter, the uncertainty is derived following the asymptotic
approximation~\cite{Cowan:2010js}. An uncertainty of 20\% is assigned to the \tttt cross section predicted by the
SM. The prediction corresponding to all parameters maximising the full likelihood is
referred to as the post-fit model.
 
The best-fit value of $\mu$ is:
\begin{equation}\nonumber
\mu=2.0 \pm 0.4\mathrm{(stat)}\,^{+0.7}_{-0.4}\mathrm{(syst)}=2.0\,^{+0.8}_{-0.6}.
\end{equation}
The systematic uncertainty is determined by subtracting in quadrature the statistical uncertainty,
obtained from a fit where all NPs are fixed to their post-fit values, from the total
uncertainty.
The measured $\mu$ value is consistent within 1.7 standard deviations with the SM
prediction corresponding to $\mu = 1$.
The probability for the background-only hypothesis to result in a signal-like
excess at least as large as seen in data
is derived using the profile-likelihood ratio following the procedure
described in Ref.~\cite{Cowan:2010js}.
From this, the significance of the observed signal is found to be 4.3 standard deviations,
while 2.4 standard deviations are expected.
Figure~\ref{fig:sr_postfit} shows the distribution of the BDT score in the signal region after performing the fit.
Good agreement is observed between data and the fitted prediction.
 
The fitted signal strength is converted into an inclusive cross section
using the SM \tttt predicted cross section of $\stttt = 12.0 \pm 2.4 ~\text{fb}$
computed at NLO in QCD and electroweak couplings~\cite{Frederix:2017wme} and excluding
its uncertainty.
The measured \tttt\ production cross section is then:
\begin{equation}\nonumber
\sigma_{\tttt}=24 \pm 5 \mathrm{(stat)}\,^{+5}_{-4}\mathrm{(syst)}\,\text{fb}=24\,^{+7}_{-6}\,\text{fb}.
\end{equation}
 
The normalisation factors of the different background sources determined from the fit
are shown in Table~\ref{tab:SSML_CRfit}. The post-fit background and signal yields
are shown in Table~\ref{tab:ssml:yields}.
The normalisation factor for the \ttW background is compatible with the observation
from the previous ATLAS $t\bar{t}H$ search~\cite{ATLAS-CONF-2019-045} where the
reference theoretical \ttW background cross section was scaled up by 20\% to account
for extra jet production and EW effects compared to the theoretical cross section used
in this analysis.
The post-fit value of the nuisance parameter associated with the systematic uncertainty
in the $t\bar{t}W$ background with $\Nj=7$ is $0.18^{+0.73}_{-0.61}$.
This corresponds to a 22\% increase in the number of $t\bar{t}W$ events with seven jets.
The post-fit value of the nuisance parameter for the systematic uncertainty of the \ttW background with $\Nj\ge8$ is $0.22^{+0.56}_{-0.42}$,
corresponding to a 65\% increase in the number of $t\bar{t}W$ events with eight or more jets.
As a result of these increases and of the change in the $t\bar{t}W$ background normalisation
factor \NFttW, the overall $t\bar{t}W$ background yield in the signal-enriched region
with a BDT score above zero increased from the $12.4 \pm 8.8$ events predicted to
$23.2 \pm 10.1$ events after the fit to data.
Apart from the uncertainties discussed above, no other nuisance parameters are found to be
significantly adjusted or constrained by the fit.
 
Figure~\ref{fig:cr_postfit} shows the yields or the discriminating variable
distributions used in the fit in each CR.
Good agreement is observed between data and the post-fit computation.
 
\begin{table}[!h]
\renewcommand*{\arraystretch}{1.4}
\centering
\caption{Normalisation factors for various backgrounds determined from the fit
to the control regions. The uncertainties include both the statistical and systematic uncertainties.}
\begin{tabular}{l|ccccc}
\hline
Parameter & \NFttW &\NFCO &\NFGst & \NFHFe & \NFHFm\\
\hline
Value & $1.6 \pm 0.3$ & $1.6 \pm 0.5$ & $0.9 \pm 0.4$ & $0.8 \pm 0.4 $ & $1.0 \pm 0.4$ \\
\hline
\end{tabular}
\label{tab:SSML_CRfit}
\end{table}
 
\begin{figure}[!htb]
\centering
\includegraphics[width=0.45\textwidth]{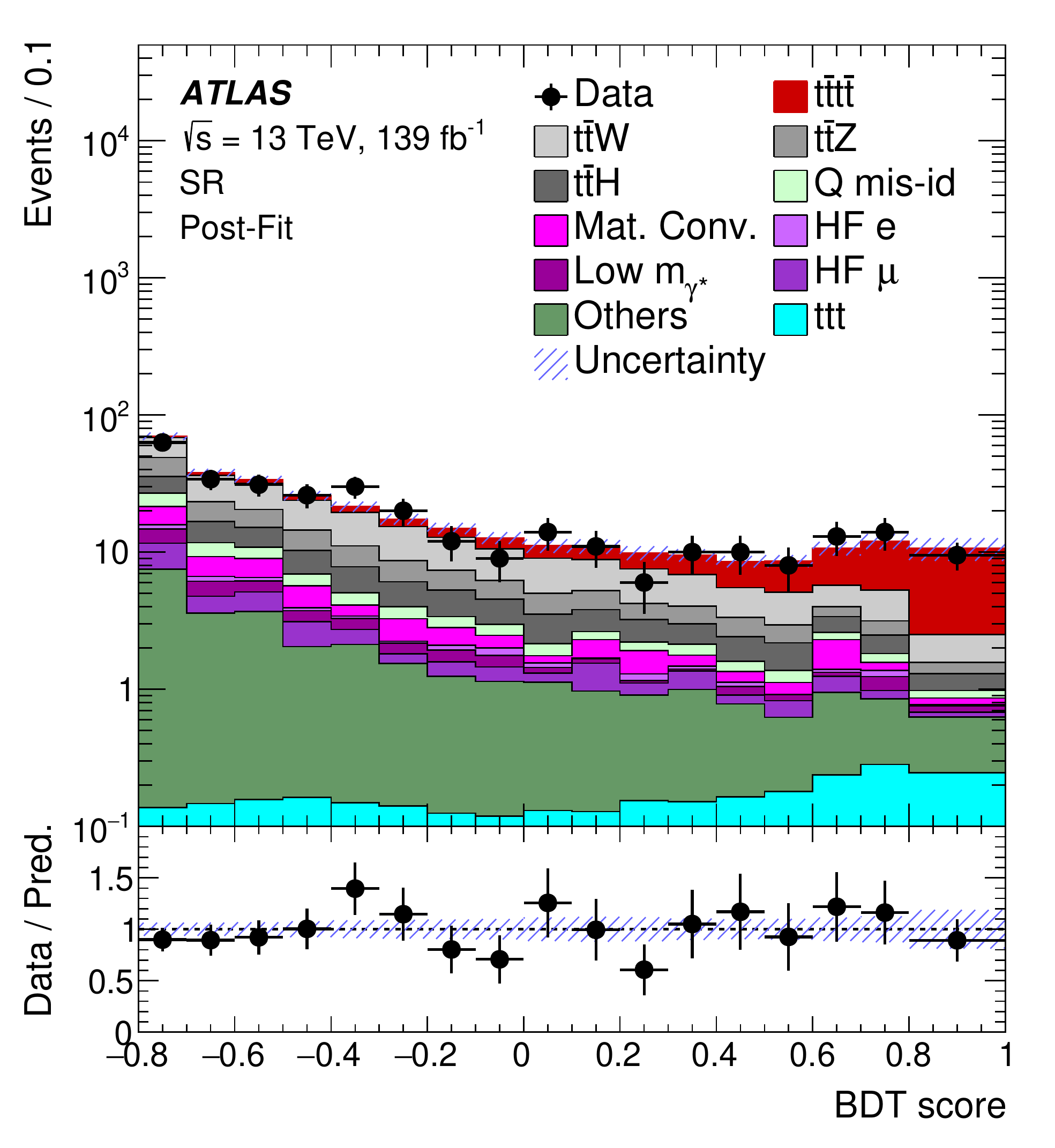}
\caption{
Comparison between data and prediction after the fit (`Post-Fit') for the distribution of the BDT score in the SR.
The band includes the total uncertainty of the post-fit computation.
The ratio of the data to the total post-fit computation is shown in the lower panel.
See Section~\ref{sec:backSSML} for the definitions of the different background categories.
}
\label{fig:sr_postfit}
\end{figure}

\begin{table}[htbp]
\begin{center}
\caption{Post-fit background and signal yields in the full signal region as well as for events in which the BDT score is also greater than zero.
The total systematic uncertainty differs from the sum in quadrature of the different uncertainties due to correlations.
Q mis-id refers to the charge misassignment background. Mat.\ Conv.
and Low $m_{\gamma^*}$ refer respectively to events with one non-prompt
electron originating from photon conversion in the detector material
and to events with a virtual photon leading to an \ee pair.
HF~$e$ (HF~$\mu$) refers to events with one non-prompt electron (muon)
from heavy-flavour hadron decay, LF refers to events with
a lepton originating from light-meson decay,
and 'Other $t\bar{t}X$' includes events coming from \ttWZ, \ttZZ, \ttWH, \ttHH.
}
\begin{tabular}{lcc}
\toprule
& {SR} & SR and BDT\,>\,0 \\
\midrule
\ttW   & 102   $\pm$ 26 & 23 $\pm$ 10 \\
\ttWW  & 7     $\pm$ 4  & 2  $\pm$ 1  \\
\ttZ   & 48    $\pm$ 9  & 9  $\pm$ 2  \\
\ttH   & 38    $\pm$ 9  & 8  $\pm$ 2  \\
Q mis-id   & 16  $\pm$ 1  & 2.7 $\pm$ 0.2 \\
Mat.\ Conv.    & 19  $\pm$ 6  & 3  $\pm$ 1 \\
Low $m_{\gamma^*}$   & 9 $\pm$ 4 & 0.9 $\pm$ 0.5 \\
HF~$e$     & 3 $\pm$ 3  & 1 $\pm$ 1 \\
HF~$\mu$   & 12 $\pm$ 6 & 3 $\pm$ 2 \\
LF & 4 $\pm$ 5  & 1 $\pm$ 1 \\
Other fake    & 6 $\pm$ 2 & 2 $\pm$ 1 \\
VV,VVV,VH & 3 $\pm$ 2 & 0.2 $\pm$ 0.2 \\
\tZq, \tWZ & 5 $\pm$ 2 & 1.0 $\pm$ 0.4\\
Other $t\bar{t}X$ & 3 $\pm$ 2 & 1 $\pm$ 1 \\
\ttt     & 3 $\pm$ 3 & 2 $\pm$ 2 \\
\midrule
Total bkg & 278 $\pm$ 22 & 59 $\pm$ 10 \\
\midrule
\tttt  & 60    $\pm$ 17 & 44 $\pm$ 12 \\
\midrule
Total  & 337 $\pm$ 18 & 103 $\pm$ 10\\
\midrule
Data   & 330 & 105 \\
\bottomrule
\end{tabular}
\label{tab:ssml:yields}
\end{center}
\end{table}
 
\begin{figure}[!htb]
\centering
\includegraphics[width=0.45\textwidth]{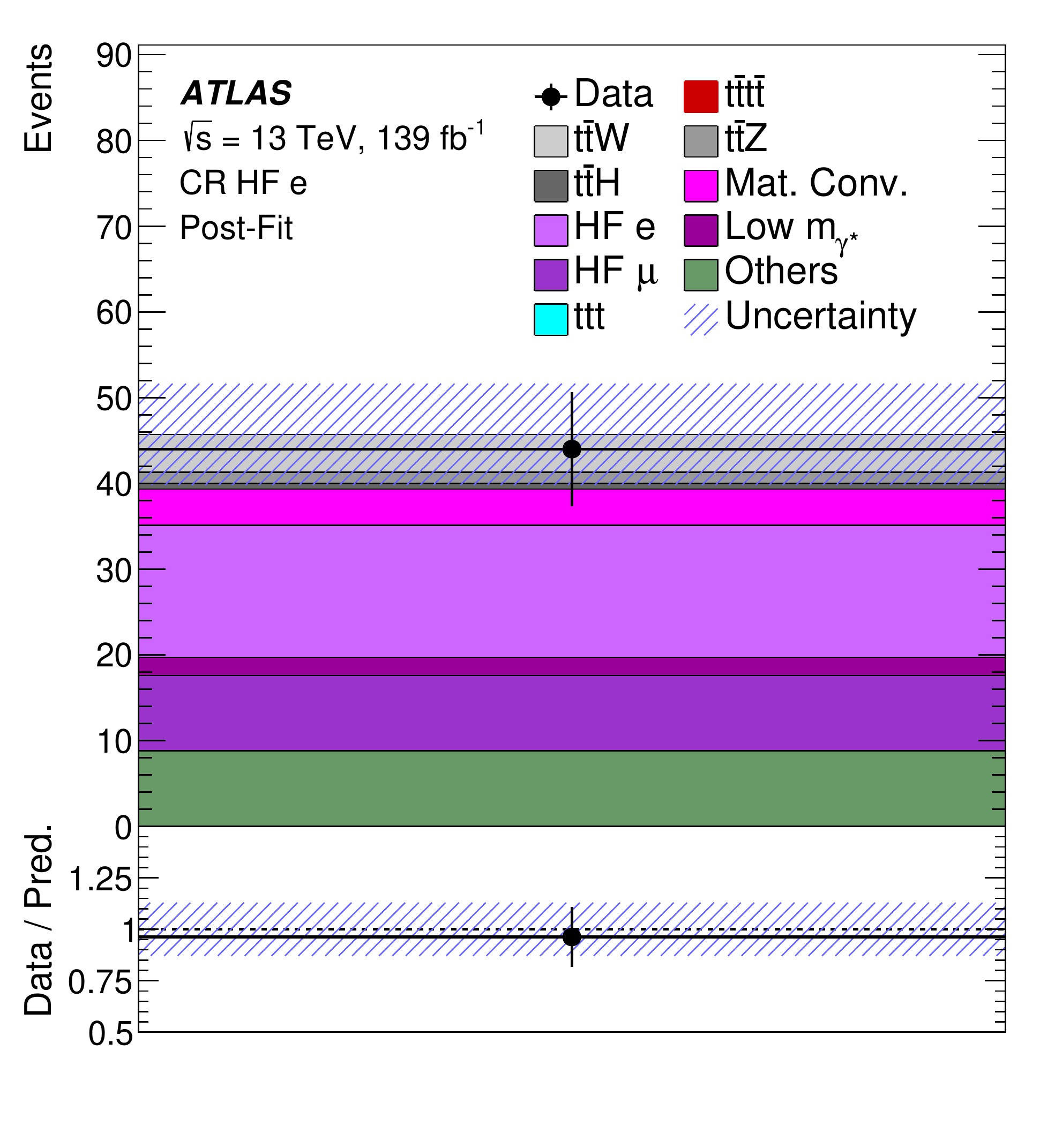}
\includegraphics[width=0.45\textwidth]{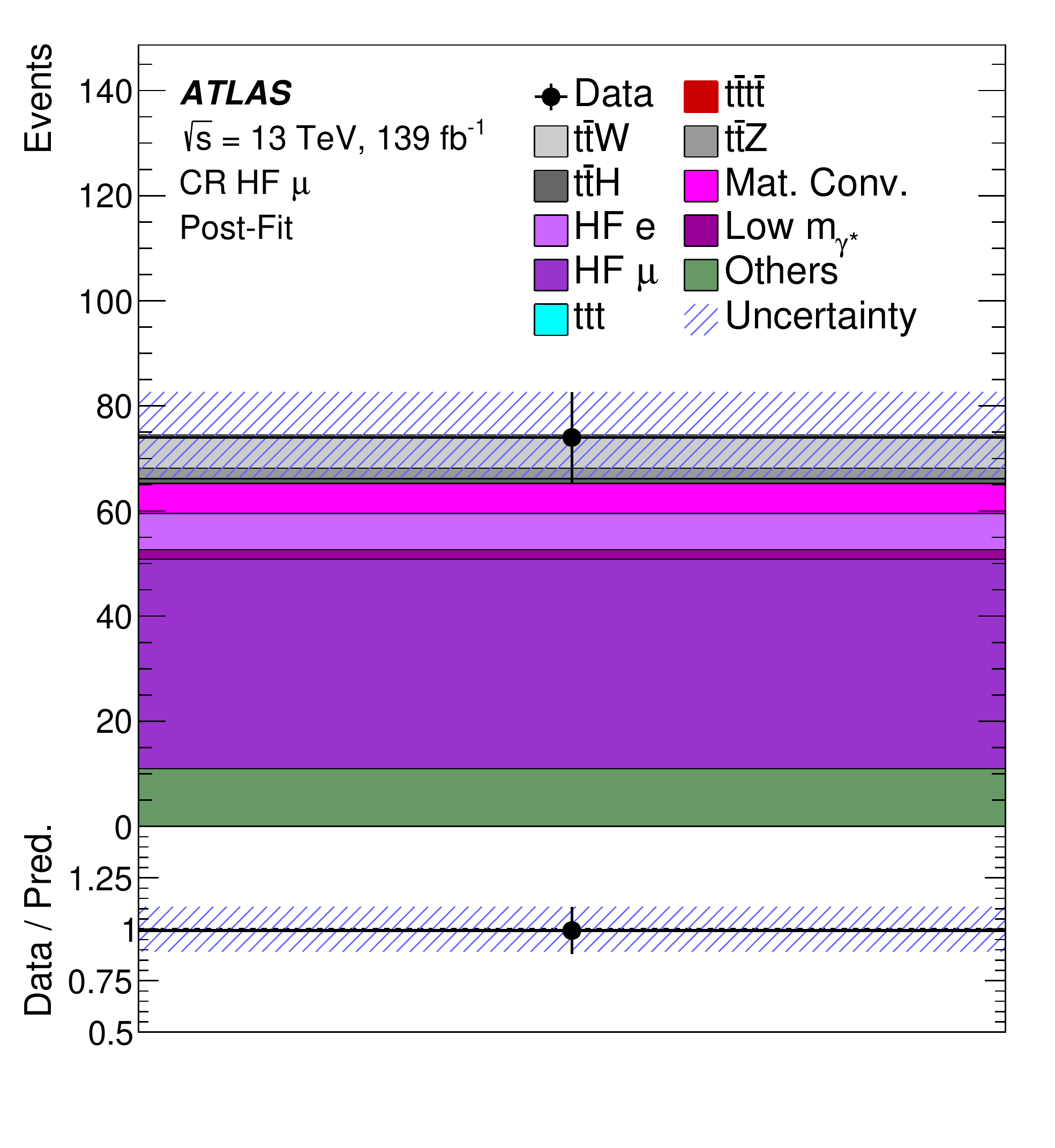}\\
\includegraphics[width=0.45\textwidth]{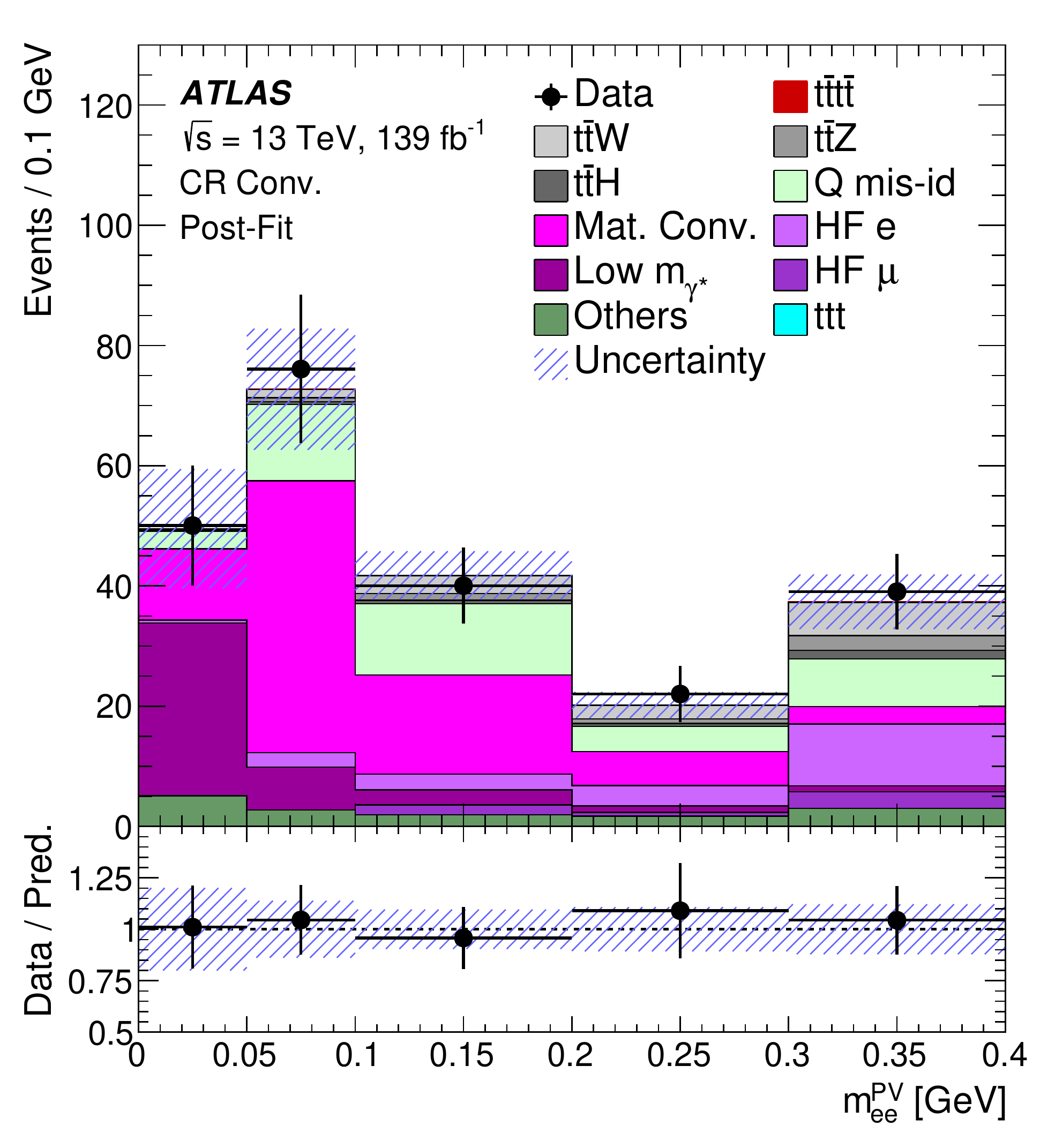}
\includegraphics[width=0.45\textwidth]{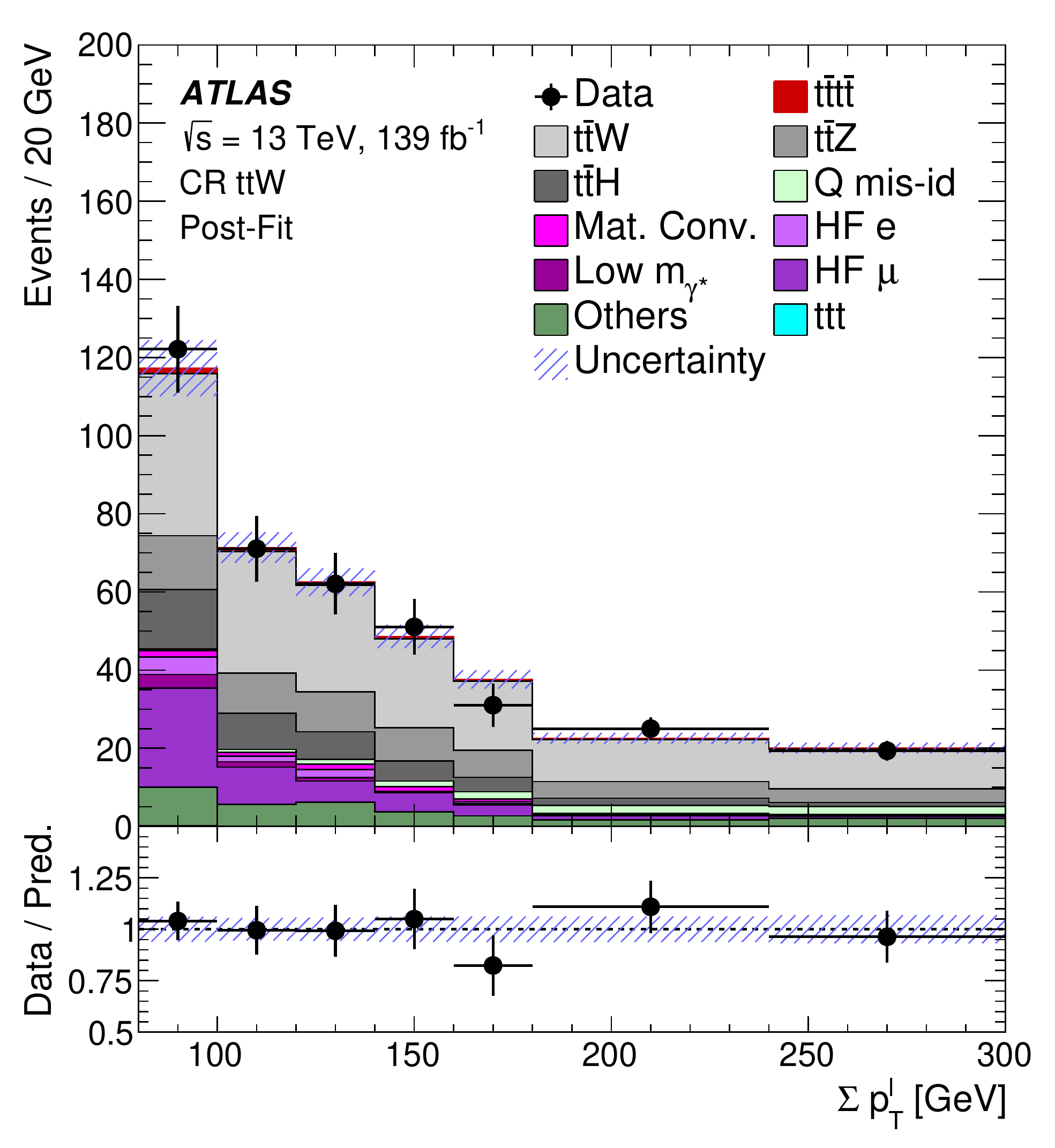}
\caption{
Comparison between data and prediction after the fit (`Post-Fit') for the
yields or distributions of
the discriminating variables used in the fit in each CR (see Table~\ref{tab:CRSR_SSML}).
The band includes the total uncertainty of the post-fit computation.
The ratio of the data to the total post-fit computation is shown in the lower panel.
The first and last bins contain underflow and overflow events, respectively.
See Section~\ref{sec:backSSML} for the definitions of the different background categories.
}
\label{fig:cr_postfit}
\end{figure}

In order to check the \ttW background normalisation and modelling,
a validation region is defined, based on the fact that the \ttW process is
charge asymmetric.
The difference between the number of events with a positive sum
and the number of events with a negative sum of the charges of the selected leptons
is built in the region with at least four jets with at least
two being $b$-tagged.
This procedure removes the charge-symmetric processes and
allows construction of distributions where \ttW events dominate.
The jet multiplicity and the BDT score distributions are displayed
in Figure~\ref{fig:ttwcr_postfit} and show good agreement
between data and post-fit computations.
 
\begin{figure}[!htb]
\centering
\includegraphics[width=0.45\textwidth]{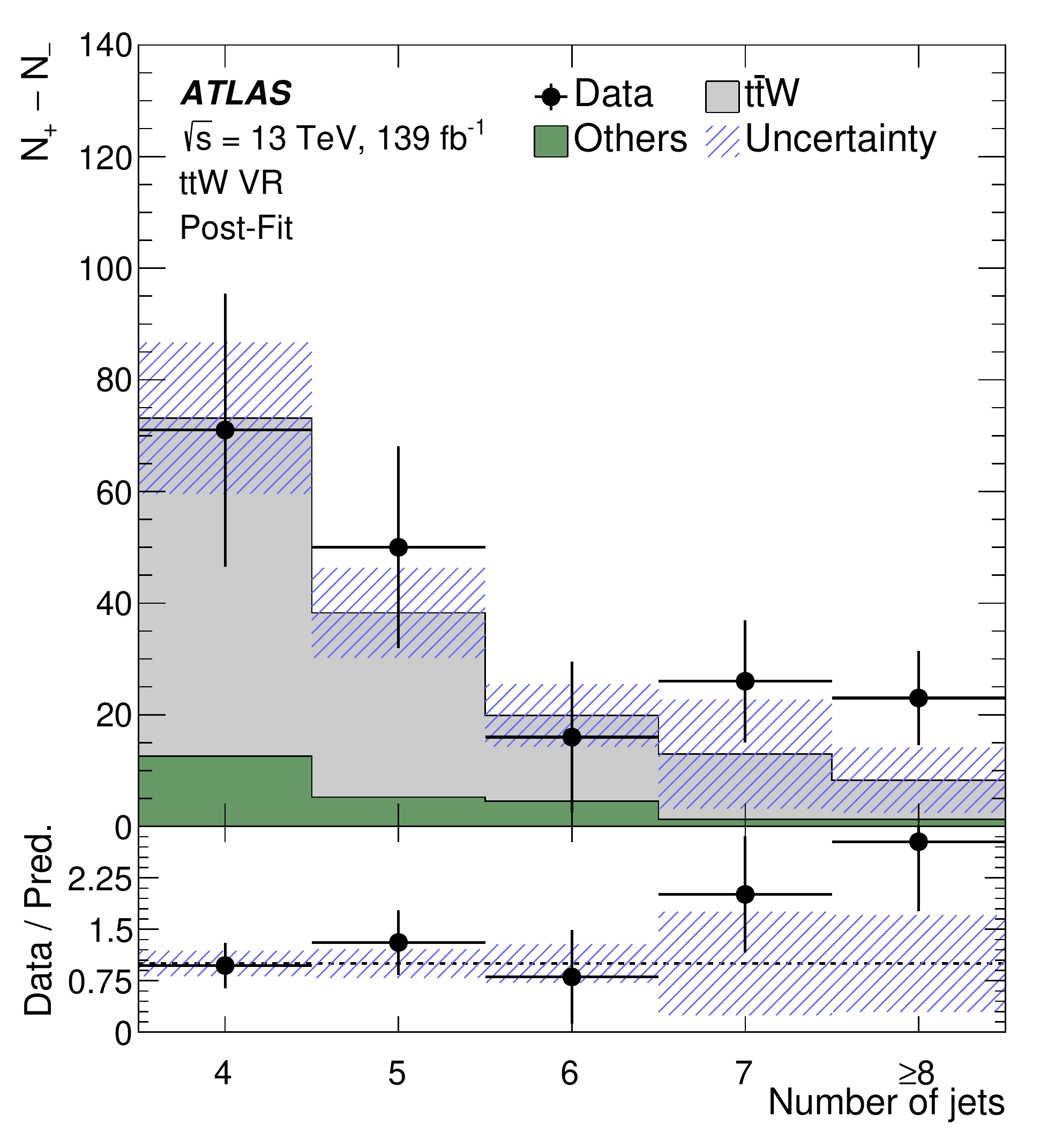}
\includegraphics[width=0.45\textwidth]{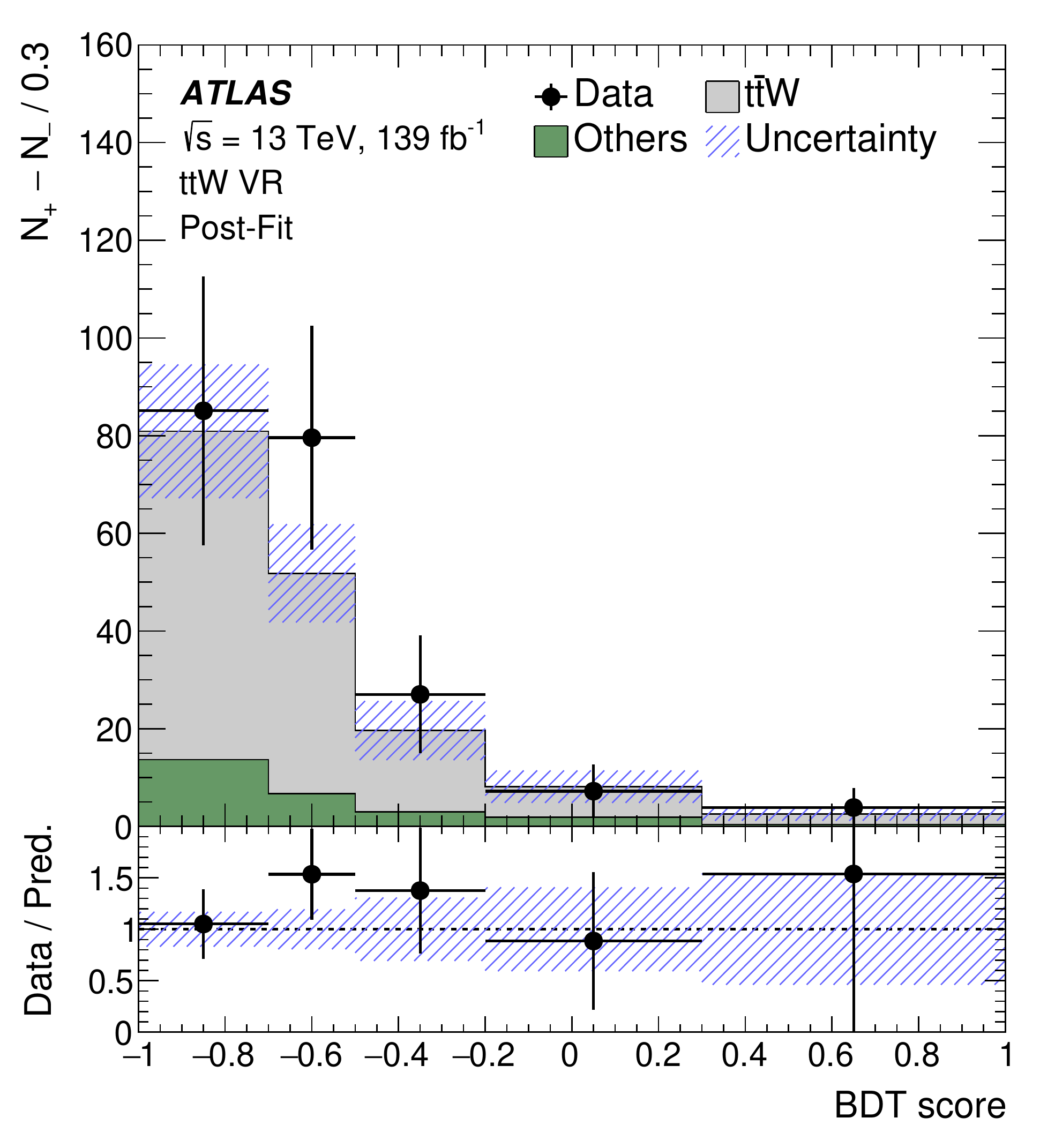}
\caption{
Post-fit comparison between data and prediction in the \ttW validation region for the multiplicity of jets (left) and the BDT score (right).
The $y$-axis label $N_{+} - N_{-}$ represents the difference between the number of events with a positive sum
and the number of events with a negative sum of the charges of the selected leptons.
The band includes the total uncertainty of the post-fit computation.
The ratio of the data to the total post-fit computation is shown in the lower panel.
The first and last bins contain underflow and overflow events, respectively.
}
\label{fig:ttwcr_postfit}
\end{figure}
 
The distributions for some of the key analysis variables are shown in
Figure~\ref{fig:varpostfit} for the events in the signal region and in Figure~\ref{fig:bdt0vars} for events
in a signal-enriched region with a BDT score above zero.
 
\begin{figure}[!htb]
\centering
\includegraphics[width=0.45\textwidth]{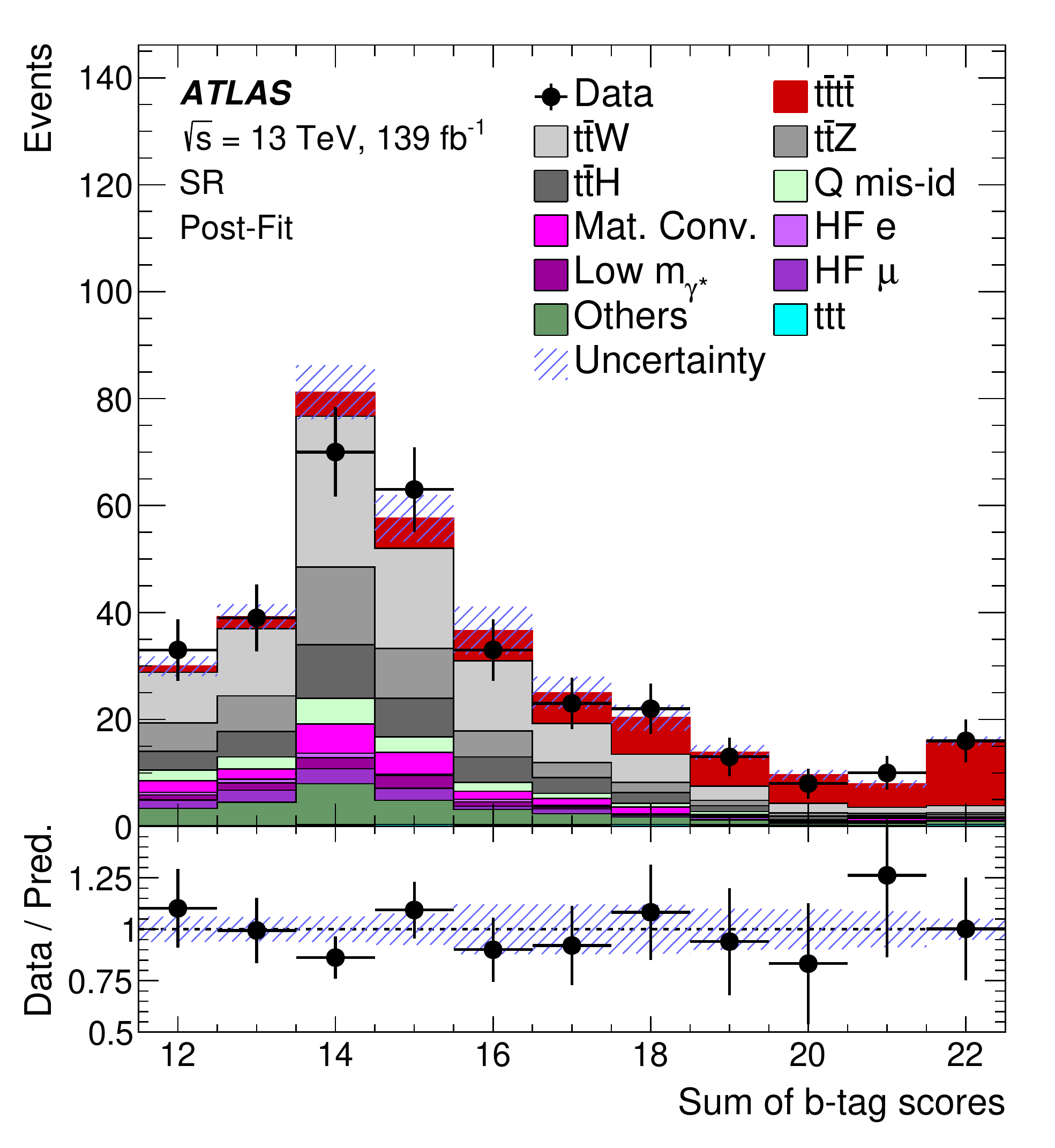}
\includegraphics[width=0.45\textwidth]{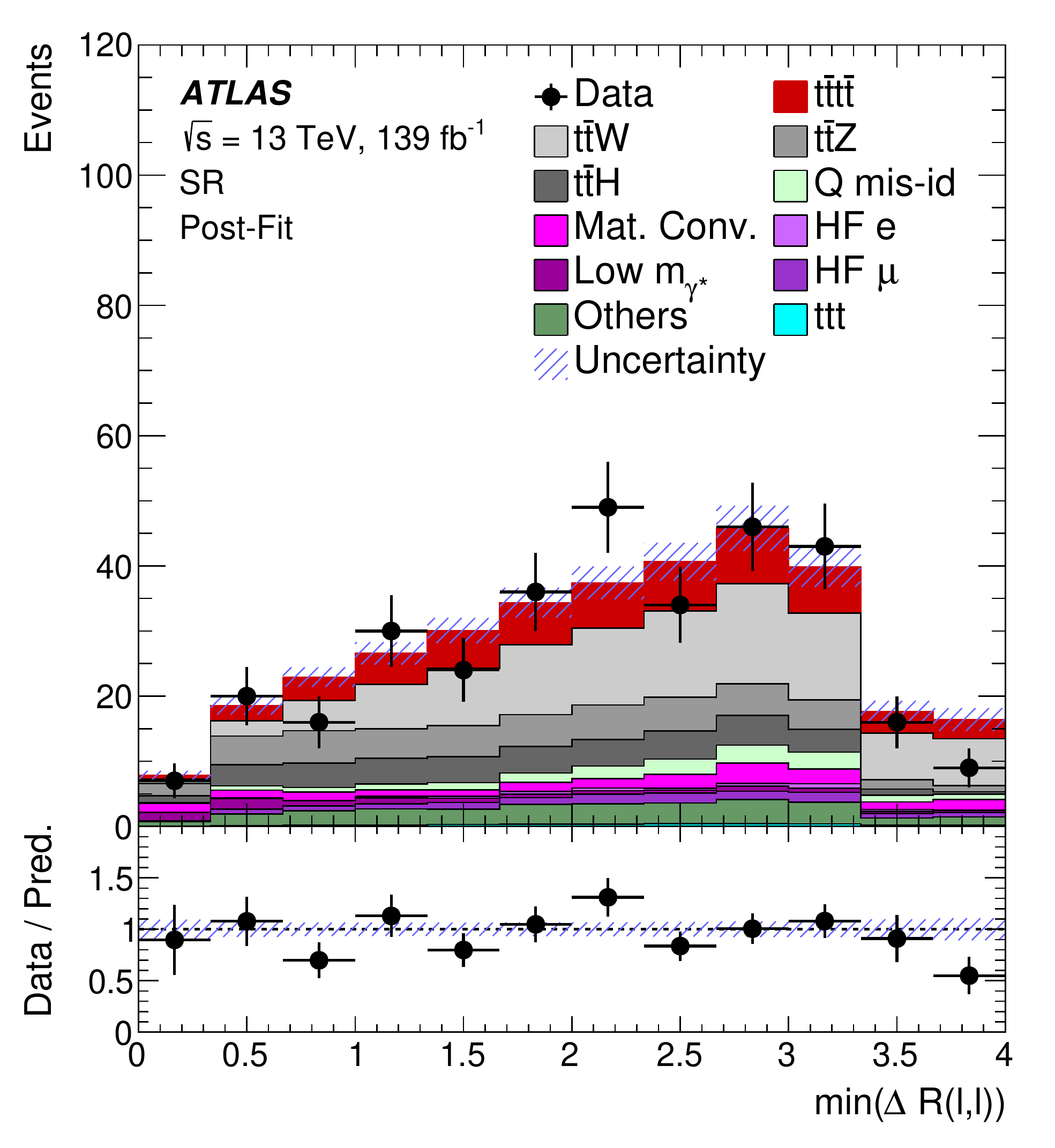}
\includegraphics[width=0.45\textwidth]{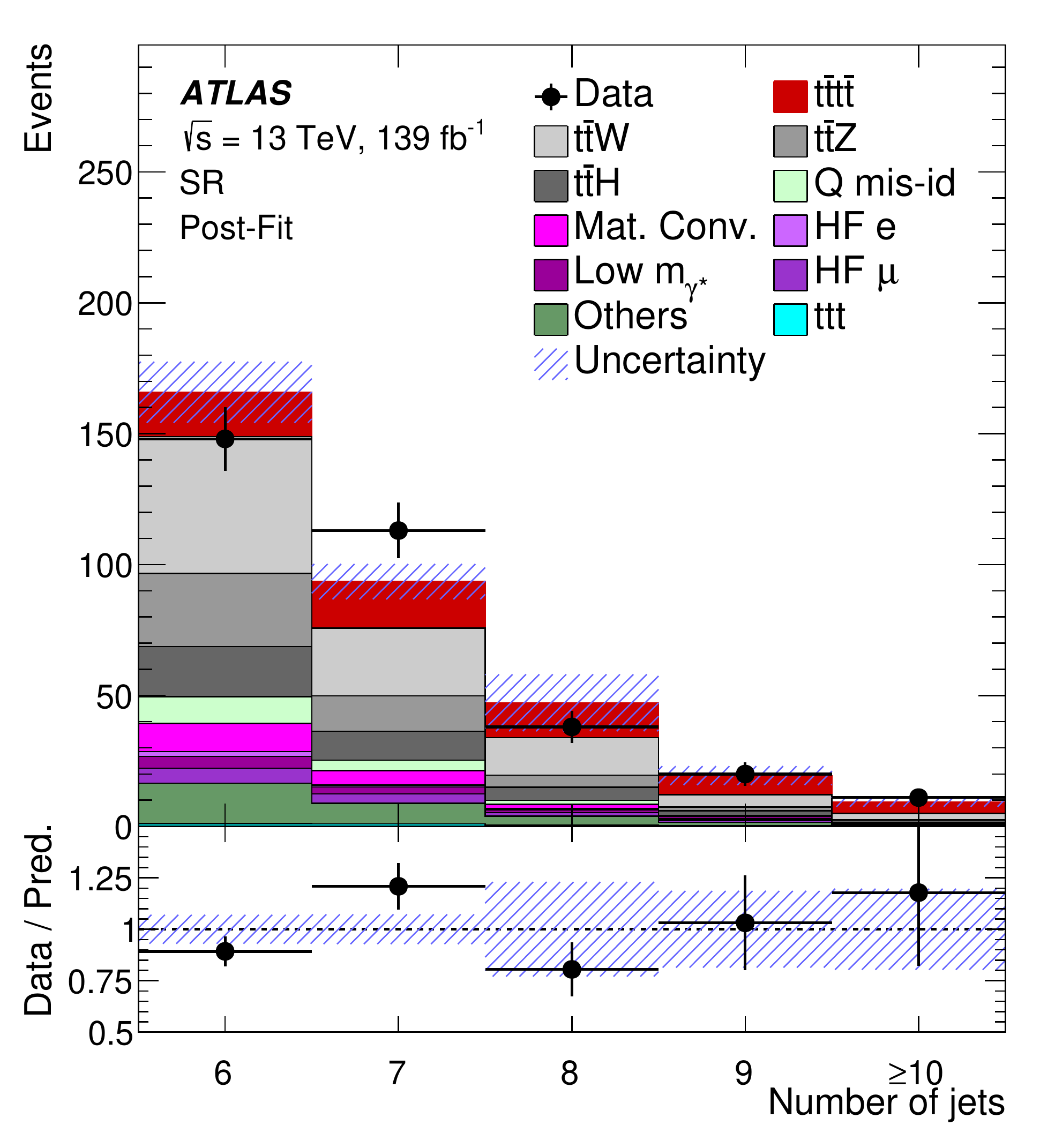}
\includegraphics[width=0.45\textwidth]{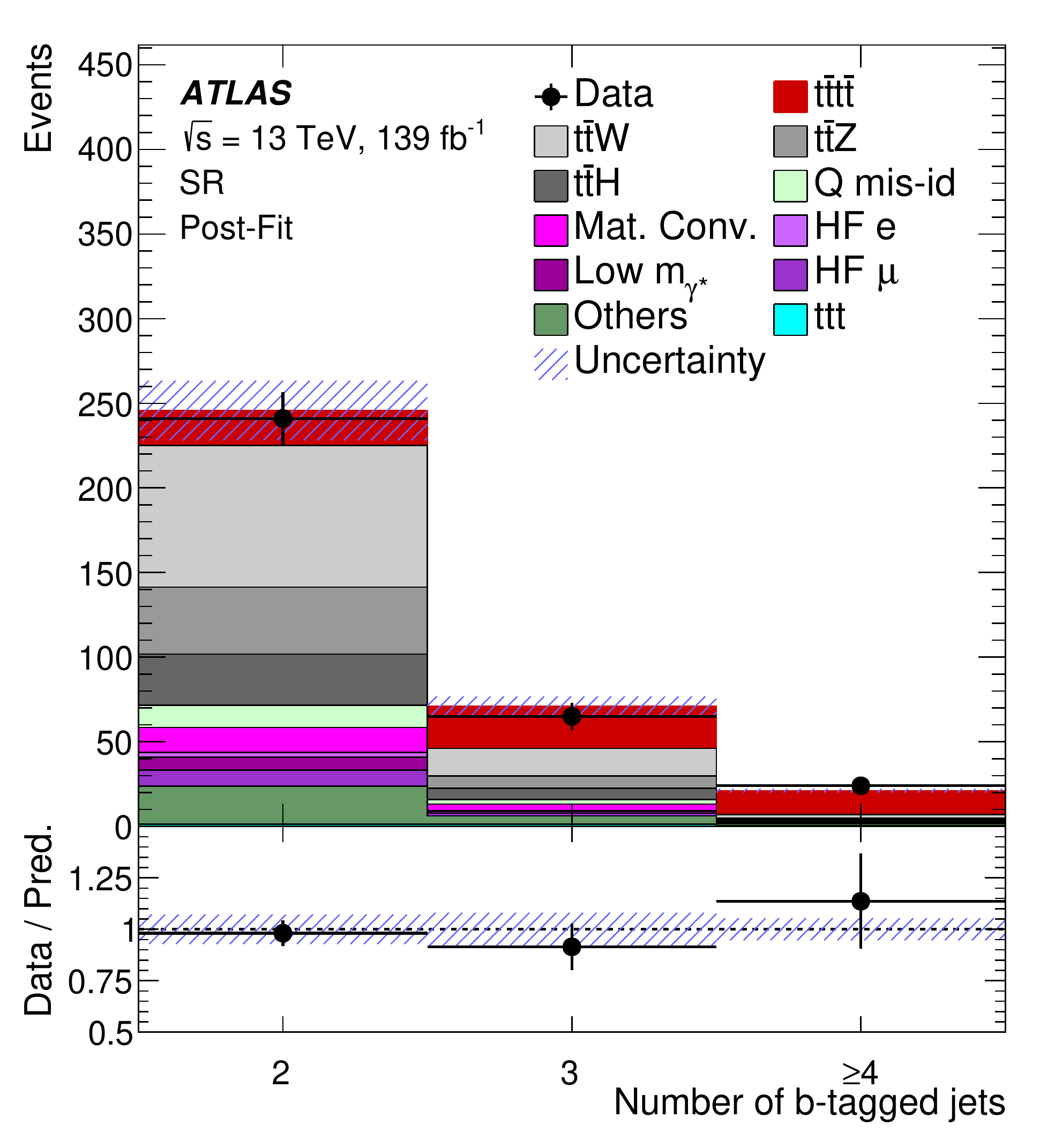}
\caption{Post-fit comparison between data and prediction for signal region events
for the distributions of: the sum of $b$-tagging pseudo-continuous scores of the jets
in the event (top left), the minimum distance between two leptons among all possible pairs (top right),
the multiplicity of jets (bottom left) and the multiplicity of $b$-tag jets (bottom right).
The band includes the total uncertainty of the post-fit computation.
The ratio of the data to the total post-fit computation is shown in the lower panel.
The first and last bins contain underflow and overflow events, respectively.
See Section~\ref{sec:backSSML} for the definitions of the different background categories.}
\label{fig:varpostfit}
\end{figure}
 
\FloatBarrier
 
\begin{figure}[!htb]
\centering
\includegraphics[width=0.45\textwidth]{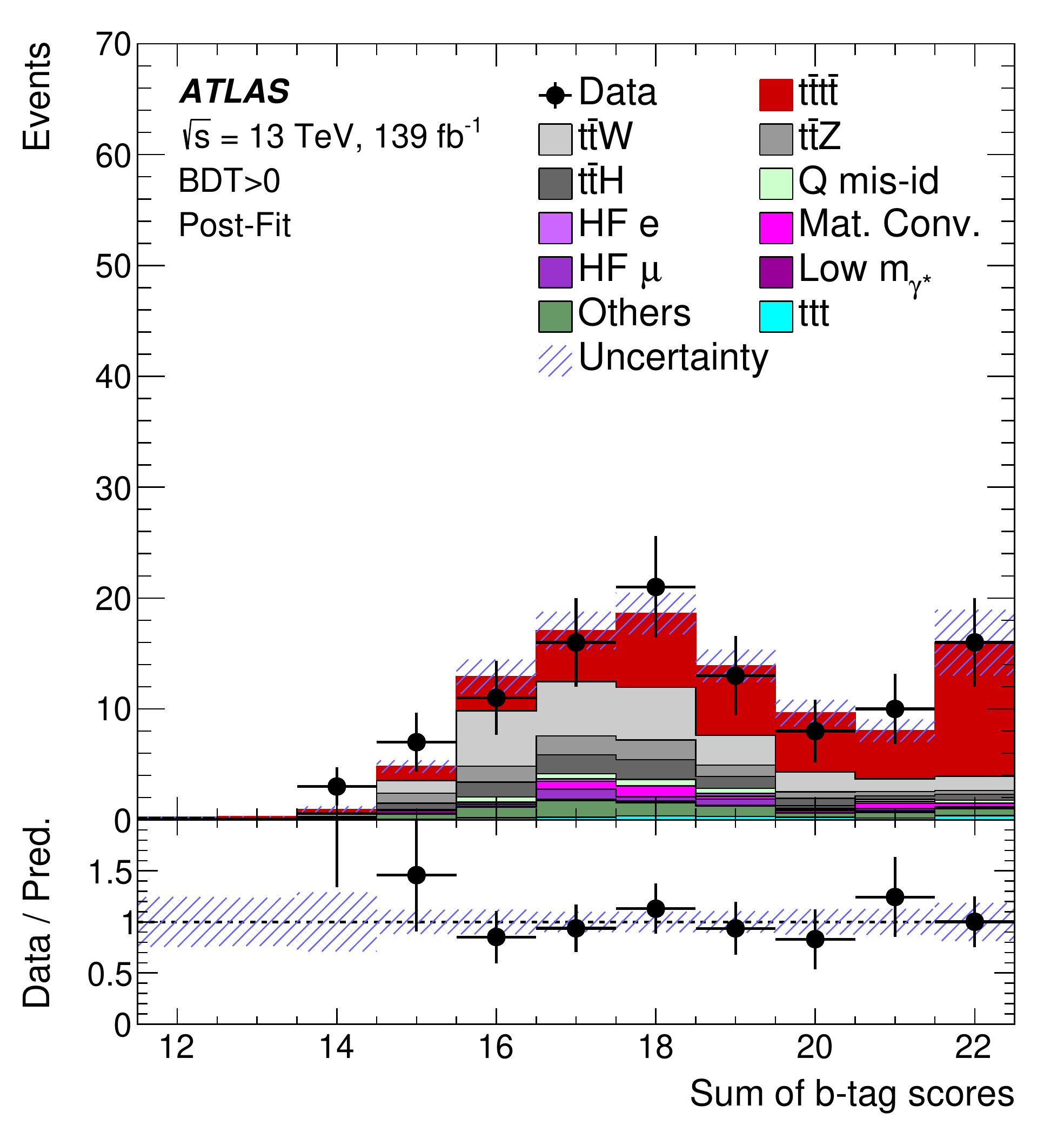}
\includegraphics[width=0.45\textwidth]{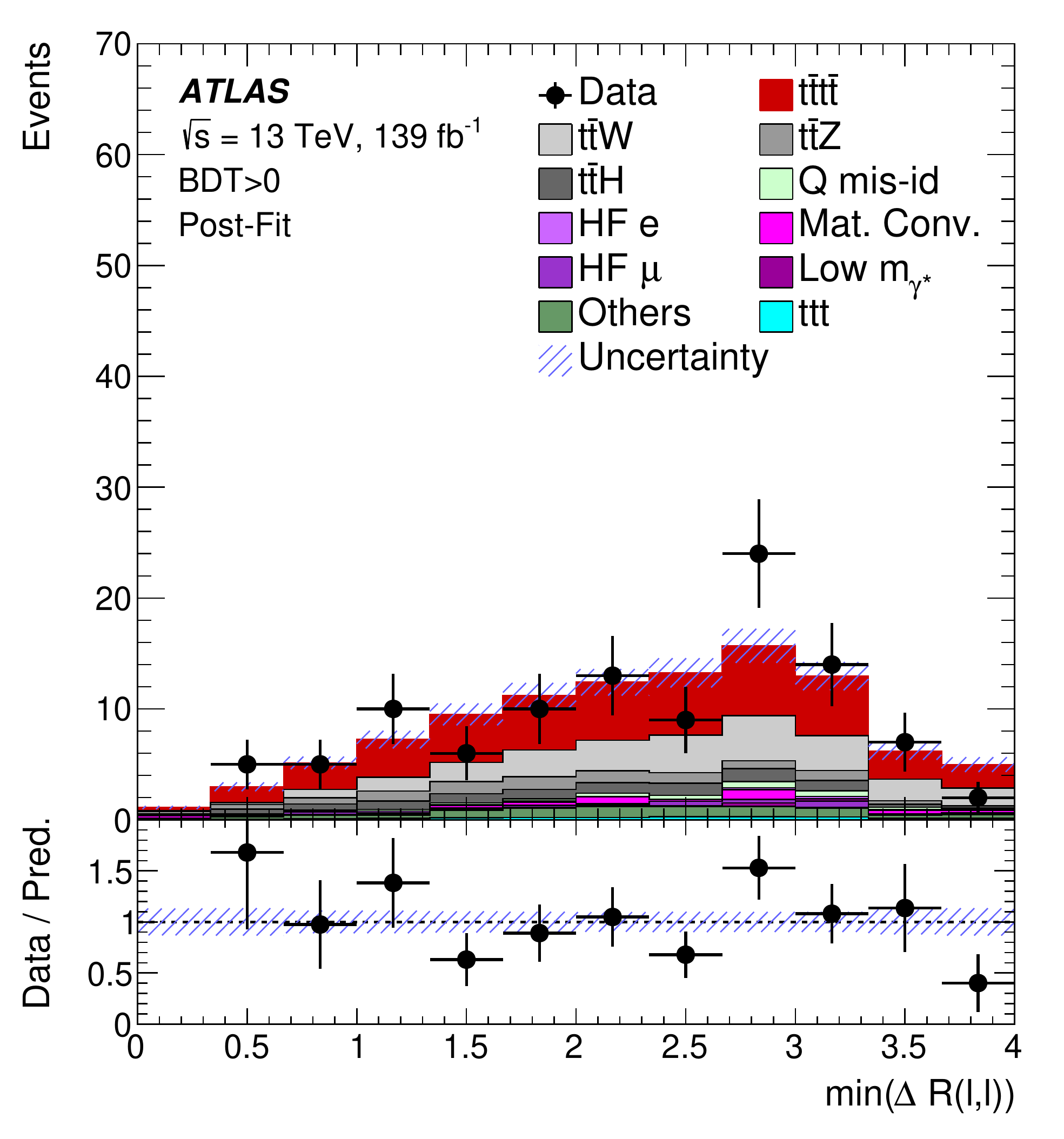}
\includegraphics[width=0.45\textwidth]{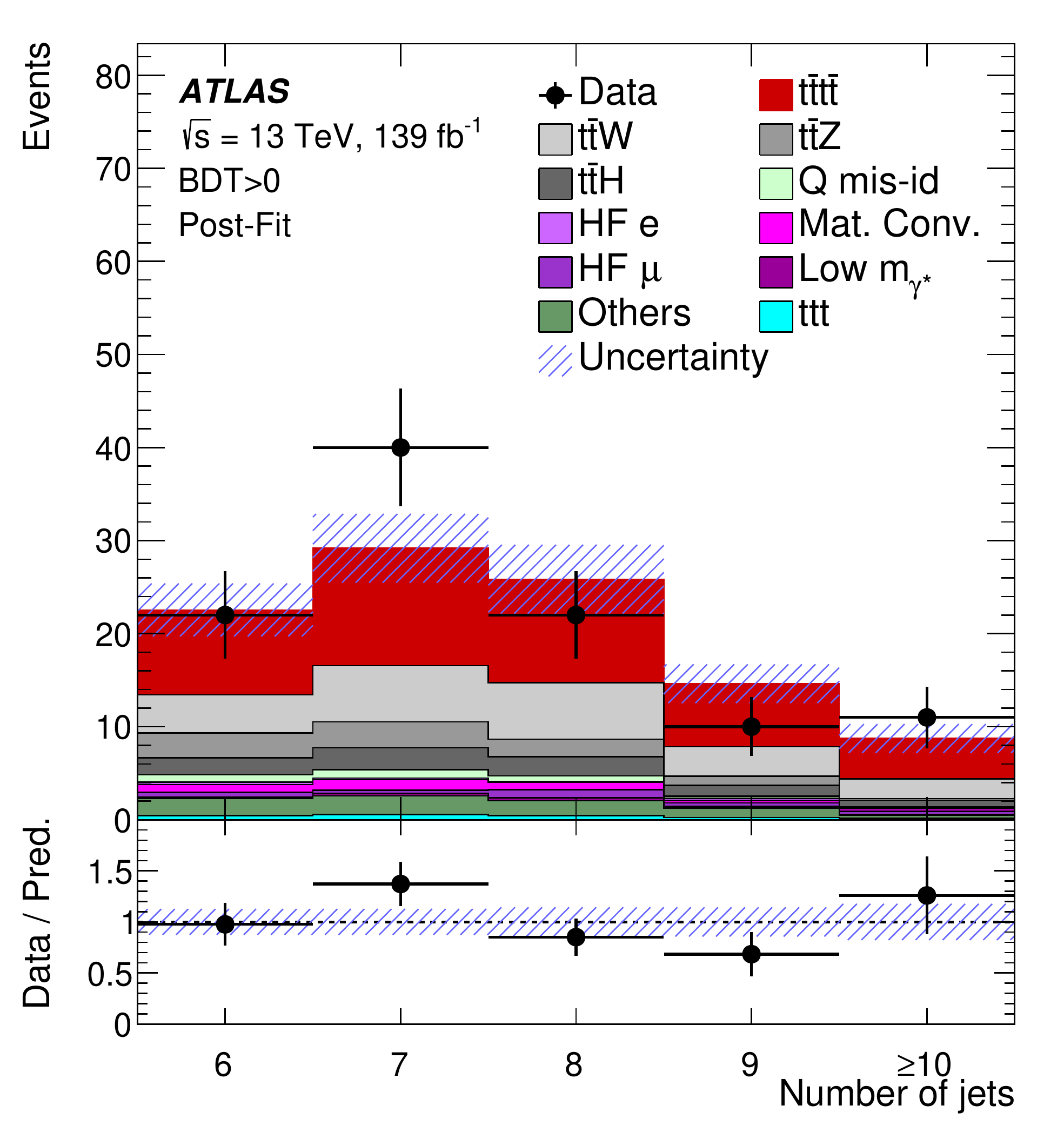}
\includegraphics[width=0.45\textwidth]{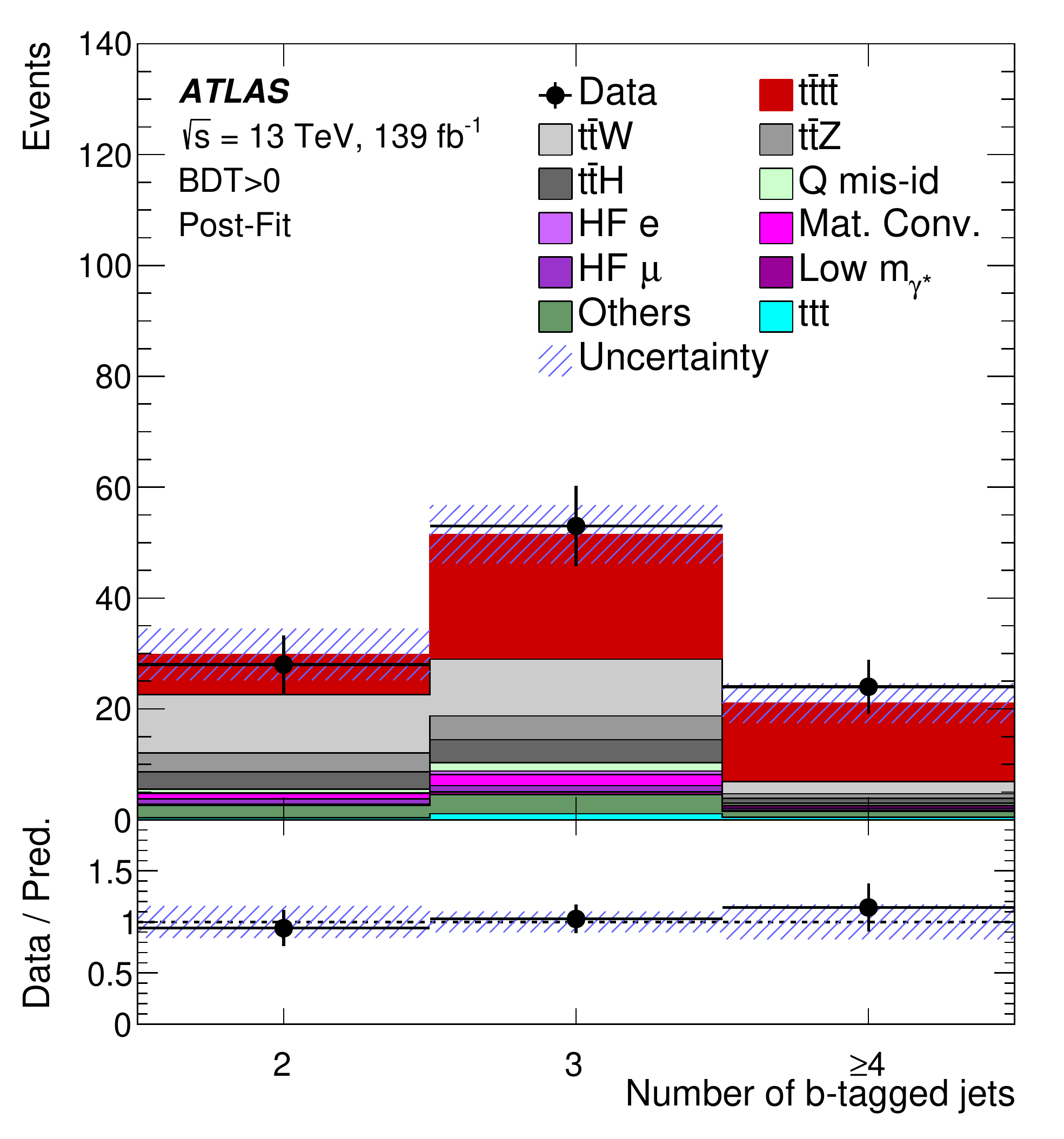}
\caption{Post-fit comparison between data and prediction for signal region events
with a BDT score greater than zero for the distributions of: the sum of $b$-tagging pseudo-continuous
scores of the jets in the event (top left), the minimum distance between two leptons among all possible pairs (top right),
the multiplicity of jets (bottom left) and the multiplicity of $b$-tag jets (bottom right).
The band includes the total uncertainty of the post-fit computation.
The ratio of the data to the total post-fit computation is shown in the lower panel.
The first and last bins contain underflow and overflow events, respectively.
See Section~\ref{sec:backSSML} for the definitions of the different background categories.}
\label{fig:bdt0vars}
\end{figure}

\FloatBarrier
The uncertainties impacting $\mu$ are summarised in Table~\ref{tab:syst}.
Apart from the theoretical uncertainty of the signal cross section, the largest systematic uncertainty
comes from the modelling of the \ttW process.
Within the uncertainties of the background modelling, the impact of the uncertainty
in \ttt production is also significant. The expected cross section of this process
is only of the order of 10\% of $\sigma_{\tttt}$.
However, the shape of the BDT score distribution for \ttt production is similar to the
one for signal, which leads to this uncertainty having a sizeable
impact on $\mu$.
In order to test the sensitivity of the \tttt\ measurement to the value of the \ttt cross section,
the fit was also performed assuming a \ttt cross section five times  larger than the SM cross section.
In that case the \tttt signal strength  $\mu$ decreases by 10\% while the fitted normalisation factors are mostly unaffected.
 
\begin{table}[htbp]
\begin{center}
\caption{List of the uncertainties in the signal strength $\mu$, grouped in
categories.
The quoted values are obtained by repeating the fit, fixing a set of
nuisance parameters of the sources corresponding to the considered
category, and subtracting in quadrature the resulting uncertainty
from the total uncertainty of the nominal fit presented in the last line.
The total uncertainty is different from the sum in quadrature of
the components due to correlations between nuisance parameters.
See Section~\ref{sec:systematics} for the description of the uncertainty sources.}
\begin{tabular}{lcc}
\toprule
Uncertainty source & \multicolumn{2}{c}{$\Delta \mu$} \\
\hline
\textbf{Signal modelling} & &\\
\tttt cross section                                &  $+0.56$ & $-0.31$\\
\tttt modelling                                    &  $+0.15$ & $-0.09$\\
\hline
\textbf{Background modelling} && \\
\ttW modelling                                     	&  $+0.26$ & $-0.27$\\
\ttt modelling						&  $+0.10$ & $-0.07$\\
Non-prompt leptons modelling                			&  $+0.05$ & $-0.04$\\
\ttH modelling                              		&  $+0.04$ & $-0.01$\\
\ttZ modelling                                		&  $+0.02$ & $-0.04$\\
Other background modelling                                      &  $+0.03$ & $-0.02$\\
Charge misassignment                                            &  $+0.01$ & $-0.02$\\
\hline
\textbf{Instrumental} &&\\
Jet uncertainties                                               &  $+0.12$ & $-0.08$\\
Jet flavour tagging (light-flavour jets)           	        &  $+0.11$ & $-0.06$\\
Simulation sample size                                   	&  $+0.06$ & $-0.06$\\
Luminosity                                                      &  $+0.05$ & $-0.03$\\
Jet flavour tagging ($b$-jets)          			&  $+0.04$ & $-0.02$\\
Jet flavour tagging ($c$-jets)                                  &  $+0.03$ & $-0.01$\\
Other experimental uncertainties                                &  $+0.03$ & $-0.01$\\
\hline
Total systematic uncertainty                                    & $+0.70$ & $-0.44$\\
\bottomrule
\textbf{Statistical}                                                  &  $+0.42$ & $-0.39$\\
Non-prompt leptons normalisation (HF, Mat.\ Conv., Low $m_{\gamma^*}$) &  $+0.05$ & $-0.04$\\
$t\bar{t}W$ normalisation                                    &  $+0.04$ & $-0.04$\\
\bottomrule
Total uncertainty                                            & $+0.83$ & $-0.60$\\
\bottomrule
\end{tabular}
\label{tab:syst}
\end{center}
\end{table}
 

The stability of the result has been checked.
The fit was repeated with the data split according to year or by splitting the signal region into two regions
with either same-sign dilepton events or events with at least three leptons.
Different fits were also performed by using only positively charged same-sign lepton pairs or only negatively charged same-sign lepton pairs.
All these tests showed compatible $\mu$ values.

An additional test was performed by splitting the SR into five regions according to the number of leptons
and $b$-tagged jets and by fitting the \HT distribution in each region.
The BDT score is therefore not used in this test.
The observed (expected) significance is found to be $4.3$ ($2.1$) and
the fitted signal strength is $2.2^{+0.9}_{-0.6}$.
This result is consistent with the result from the default fit.

 
\FloatBarrier
 
 
\section{Conclusion}
 
A search is presented for four-top-quark production using an integrated luminosity of \lumi~\ifb\
of proton--proton collision data at $\sqrt{s}=13~\tev$ collected by the ATLAS detector
at the LHC.
The observed (expected)
significance with respect to the background-only hypothesis is
4.3 (2.4) standard deviations, providing evidence for this process
assuming the Standard Model four-top-quark production properties.
The four-top-quark production cross section is measured to
be $24^{+7}_{-6}$~fb which
is consistent within 1.7 standard deviations with
the Standard Model expectation of $\stttt = 12.0 \pm 2.4 ~\text{fb}$.

\section*{Acknowledgements}
 

We thank CERN for the very successful operation of the LHC, as well as the
support staff from our institutions without whom ATLAS could not be
operated efficiently.
 
We acknowledge the support of ANPCyT, Argentina; YerPhI, Armenia; ARC, Australia; BMWFW and FWF, Austria; ANAS, Azerbaijan; SSTC, Belarus; CNPq and FAPESP, Brazil; NSERC, NRC and CFI, Canada; CERN; ANID, Chile; CAS, MOST and NSFC, China; COLCIENCIAS, Colombia; MSMT CR, MPO CR and VSC CR, Czech Republic; DNRF and DNSRC, Denmark; IN2P3-CNRS and CEA-DRF/IRFU, France; SRNSFG, Georgia; BMBF, HGF and MPG, Germany; GSRT, Greece; RGC and Hong Kong SAR, China; ISF and Benoziyo Center, Israel; INFN, Italy; MEXT and JSPS, Japan; CNRST, Morocco; NWO, Netherlands; RCN, Norway; MNiSW and NCN, Poland; FCT, Portugal; MNE/IFA, Romania; JINR; MES of Russia and NRC KI, Russian Federation; MESTD, Serbia; MSSR, Slovakia; ARRS and MIZ\v{S}, Slovenia; DST/NRF, South Africa; MICINN, Spain; SRC and Wallenberg Foundation, Sweden; SERI, SNSF and Cantons of Bern and Geneva, Switzerland; MOST, Taiwan; TAEK, Turkey; STFC, United Kingdom; DOE and NSF, United States of America. In addition, individual groups and members have received support from BCKDF, CANARIE, Compute Canada, CRC and IVADO, Canada; Beijing Municipal Science \& Technology Commission, China; COST, ERC, ERDF, Horizon 2020 and Marie Sk{\l}odowska-Curie Actions, European Union; Investissements d'Avenir Labex, Investissements d'Avenir Idex and ANR, France; DFG and AvH Foundation, Germany; Herakleitos, Thales and Aristeia programmes co-financed by EU-ESF and the Greek NSRF, Greece; BSF-NSF and GIF, Israel; La Caixa Banking Foundation, CERCA Programme Generalitat de Catalunya and PROMETEO and GenT Programmes Generalitat Valenciana, Spain; G\"{o}ran Gustafssons Stiftelse, Sweden; The Royal Society and Leverhulme Trust, United Kingdom.
 
The crucial computing support from all WLCG partners is acknowledged gratefully, in particular from CERN, the ATLAS Tier-1 facilities at TRIUMF (Canada), NDGF (Denmark, Norway, Sweden), CC-IN2P3 (France), KIT/GridKA (Germany), INFN-CNAF (Italy), NL-T1 (Netherlands), PIC (Spain), ASGC (Taiwan), RAL (UK) and BNL (USA), the Tier-2 facilities worldwide and large non-WLCG resource providers. Major contributors of computing resources are listed in Ref.~\cite{ATL-SOFT-PUB-2020-001}.
 

\printbibliography
 
\clearpage 
 
\begin{flushleft}
\hypersetup{urlcolor=black}
{\Large The ATLAS Collaboration}

\bigskip

\AtlasOrcid[0000-0002-6665-4934]{G.~Aad}$^\textrm{\scriptsize 102}$,    
\AtlasOrcid[0000-0002-5888-2734]{B.~Abbott}$^\textrm{\scriptsize 128}$,    
\AtlasOrcid{D.C.~Abbott}$^\textrm{\scriptsize 103}$,    
\AtlasOrcid[0000-0002-2788-3822]{A.~Abed~Abud}$^\textrm{\scriptsize 36}$,    
\AtlasOrcid[0000-0002-1002-1652]{K.~Abeling}$^\textrm{\scriptsize 53}$,    
\AtlasOrcid[0000-0002-2987-4006]{D.K.~Abhayasinghe}$^\textrm{\scriptsize 94}$,    
\AtlasOrcid[0000-0002-8496-9294]{S.H.~Abidi}$^\textrm{\scriptsize 167}$,    
\AtlasOrcid[0000-0002-8279-9324]{O.S.~AbouZeid}$^\textrm{\scriptsize 40}$,    
\AtlasOrcid{N.L.~Abraham}$^\textrm{\scriptsize 156}$,    
\AtlasOrcid[0000-0001-5329-6640]{H.~Abramowicz}$^\textrm{\scriptsize 161}$
% ...
\bigskip
\\

% ...
$^{36}$CERN, Geneva; Switzerland.\\
% ...
$^{40}$Niels Bohr Institute, University of Copenhagen, Copenhagen; Denmark.\\
% ...
$^{53}$II. Physikalisches Institut, Georg-August-Universit\"{a}t G\"ottingen, G\"ottingen; Germany.\\
% ...
$^{94}$Department of Physics, Royal Holloway University of London, Egham; United Kingdom.\\
% ...
$^{102}$CPPM, Aix-Marseille Universit\'e, CNRS/IN2P3, Marseille; France.\\
$^{103}$Department of Physics, University of Massachusetts, Amherst MA; United States of America.\\
% ...
$^{128}$Homer L. Dodge Department of Physics and Astronomy, University of Oklahoma, Norman OK; United States of America.\\
% ...
$^{156}$Department of Physics and Astronomy, University of Sussex, Brighton; United Kingdom.\\
% ...
$^{161}$Raymond and Beverly Sackler School of Physics and Astronomy, Tel Aviv University, Tel Aviv; Israel.\\
% ...
$^{167}$Department of Physics, University of Toronto, Toronto ON; Canada.\\
% ...

\end{flushleft}


\end{document}